\documentclass[aps,prb,twocolumn,superscriptaddress,showpacs,floatfix]{revtex4-2}
\usepackage{comment}
\usepackage{physics}
\usepackage{amsmath,amsthm,amssymb,amsfonts, color, comment, graphicx}%, %fancyhdr, environ}
\usepackage{dsfont}
\usepackage{xcolor}
\usepackage{float}
\usepackage{graphicx}
\usepackage{dcolumn}% Align table columns on decimal point
\usepackage{bm}% bold math
\UseRawInputEncoding
\usepackage{hyperref}% add hypertext capabilities
\def\thefootnote{$^\parallel$}\footnotetext{These authors made equivalent contributions to this work.}
\usepackage[normalem]{ulem}

\newcommand{\bea}{\begin{eqnarray}}
\newcommand{\eea}{\end{eqnarray}}

\newcommand{\bS}{\mathbf{S}}

\newcommand{\br}{\mathbf{r}}

\newcommand{\bod}{\boldsymbol{\delta}}
\newcommand{\bor}{\mathbf{r}}
\newcommand{\bok}{\mathbf{k}}
\newcommand{\boq}{\mathbf{Q}}

\newcommand{\be}{\begin{equation}}
\newcommand{\ee}{\end{equation}}
\newcommand{\bk}{{{\bf{k}}}}
\newcommand{\bK}{{{\bf{K}}}}

\newcommand{\bQ}{{{\bf{Q}}}}

\newcommand{\beal}{\begin{align}}
\newcommand{\eeal}{\end{align}}

\newcommand{\pdg}{{\phantom\dagger}}

\newcommand{\dk}{{\int \frac{d^2\bk}{(2\pi)^2}}}
\newcommand{\dr}{{\int d^2\br}}

\newcommand{\btjstrw}{\mathrel{{\rotatebox[origin=c]{90}
{$\bowtie$}}\kern-0.18em\raisebox{-.95ex}{$\bullet$}
\kern-0.5em\raisebox{.97ex}{$\bullet$}
\kern-1.12em\raisebox{.97ex}{$\bullet$}
\kern-0.52em\raisebox{-.95ex}{$\bullet$}}}

\newcommand{\btjnbrR}{{\mathrel{\rotatebox[origin=c]{90}
{$\bowtie$}}\kern-0.22em\raisebox{.9ex}{$\bullet$}
\kern-1.em\raisebox{-.8ex}{$\bullet$}}}
\newcommand{\btjnbrL}{{\mathrel{\rotatebox[origin=c]{90}
{$\bowtie$}}\kern-0.22em\raisebox{-.8ex}{$\bullet$}
\kern-1.em\raisebox{+.9ex}{$\bullet$}}}

\def\a{\alpha}
\def\b{\beta}
\def\c{\chi}
\def\d{\delta}
\def\e{\epsilon}

\def\g{\gamma}

\def\n{\nu}
\def\p{\pi}

\def\s{\sigma}
\def\t{\tau}

\def\w{\omega}

\def\D{\Delta}

\def\G{\Gamma}
\def\W{\Omega}

\def\ma{{\mathcal{A}}}
\def\mc{{\mathcal{C}}}

\def\mh{{\mathcal{H}}}
\def\mi{{\mathcal{I}}}

\def\mm{{\mathcal{M}}}

\def\mt{{\mathcal{T}}}

\newcommand{\llangle}[1][]{\savebox{\@brx}{\(\m@th{#1\langle}\)}%
  \mathopen{\copy\@brx\kern-0.5\wd\@brx\usebox{\@brx}}}
\newcommand{\rrangle}[1][]{\savebox{\@brx}{\(\m@th{#1\rangle}\)}%
  \mathclose{\copy\@brx\kern-0.5\wd\@brx\usebox{\@brx}}}
\begin{document}

\preprint{APS/123-QED}

%\title{Impact of electron-electron interactions in bilayer skyrmion crystals: Symmetry breaking patterns and spontaneous topological Hall effect}
\title{Charge ordering and spontaneous topological Hall effect in bilayer skyrmion crystals}
\author{Andrew Hardy
\thefootnote{}}
\email{andrew.hardy@mail.utoronto.ca}
\author{Anjishnu Bose
\thefootnote{}}
\email{anjishnu.bose@mail.utoronto.ca}
\author{Tanmay Grover}
\author{Arun Paramekanti}
 \email{arun.paramekanti@utoronto.ca}
\affiliation{Department of Physics, University of Toronto, 60 St. George Street, Toronto, ON, M5S 1A7 Canada}

\date{\today}% It is always \today, today,

\begin{abstract}
Magnetic skyrmion crystals with zero net skyrmion charge and zero topological Hall response are interesting candidate phases which 
can occur at a vanishing magnetic field in centrosymmetric systems. We study a minimal
bilayer model of skyrmion crystals having opposite chirality and topological charge in the two layers, and show 
that it can host nearly flat electronic bands with quasi-uniform Berry curvature and quantum metric. Using Hartree-Fock
theory, we show that weak to moderate short-range electron interactions induce two distinct types of
symmetry breaking patterns depending on the band dispersion: an intra-unit-cell charge density modulation from Chern band 
mixing or a layer-imbalanced phase with a nonzero ferroelectric polarization. Both phases break inversion symmetry leading to a spontaneous and large 
net topological Hall effect, with the phase diagram tunable by external electric fields. Our results may be relevant to centrosymmetric
skyrmion materials such as Gd$_2$PdSi$_3$ and Gd$_3$Ru$_4$Al$_{12}$ as well as 
artificially engineered
heterostructures. We also discuss its relation to
recent work on twisted transition metal dichalcogenide bilayers.
\end{abstract}

\maketitle
%\section{Introduction}\label{sec:introduction}
Skyrmions were originally introduced in particle physics to describe baryons as topological field configurations (solitons) of a pion field \cite{skyrme_unified_1962,wittenQCDbook}. 
They have since been realized in many different solid state materials as  noncoplanar magnetic or electric dipole textures carrying an integer topological charge \cite{Das2019, Jiang2017, Fert2017}. 
The topology of skyrmion textures renders them stable against thermal fluctuations, making them potential candidates for magnetic 
bits and for incorporation into devices for next-generation information technologies
\cite{Wiesendanger2016, Smejkal2018}.
The topology of skyrmions also induces nontrivial real space Berry phases for electrons moving 
in skyrmion backgrounds - this Berry phases acts as an emergent $U(1)$ gauge field inducing a `topological Hall effect' (THE)  
\cite{Ye1999,Taguchi2001,Onoda2004,Bruno2004,Nagaosa2015}
Itinerant electrons moving in skyrmion crystal background form Chern bands which mimic
Landau levels, providing a band description for the THE \cite{Nagaosa2010, Nagaosa2015,  Sorn2019, Divic2021}, while
interband electronic transitions between such Chern bands 
may explain the recently observed nontrivial magneto-optical responses of skyrmion crystals \cite{Bhowmick2018,Feng2020,Sorn2021,Kato2023,Li2024}.

Magnetic skyrmions have been conventionally explored in non-centrosymmetric systems, where the breaking of inversion symmetry leads to the 
Dzyaloshinskii-Moriya (DM) exchange 
interaction. The DM interaction acts the driving force responsible for generating multiple spin spiral modes 
\cite{Rosler2006, Binz2006,Binz2006b} which coherently add up to form skyrmion crystals.
Such DM interactions lead to skyrmion lattices in bulk three-dimensional (3D) crystals such as MnSi \cite{Muhlbauer2009,Lee2009} and
Fe$_{1-x}$Co$_x$Si \cite{Munzer2010} or at 2D interfaces between different
materials where inversion breaking can generate a large interfacial DM interaction \cite{Banerjee2014, Rowland2016, Matsuno2016,Caretta2020,Wu2022}. \par

In contrast to the above systems, skyrmions can also appear in collinear antiferromagnets (AFs) where the N\'eel vector forms
nontrivial spin textures. We may think of these as arising from spins on two sublattices, subject to the same DM interaction, 
forming their own topological
textures, with the two sublattices being locked antiparallel to each other \cite{Gobel2017a}. They may also form with a bilayer degree of freedom replacing the sublattice \cite{ Zhang2016b}. Such an antiferro-skyrmion crystal (AF-SkX) 
will exhibit a vanishing THE but can exhibit a nontrivial spin-THE, reminiscent of the spin Hall effect in the Kane-Mele model
\cite{Barker2016, Gobel2017a, Akosa2018, Kane2005}. The experimental observation of topological spin textures in AFs is relatively recent, including a fractional AF-SkX in MnSc$_2$S$_4$ \cite{Gao2020},  AF half-skyrmions and bimerons in 
$\alpha$-Fe$_2$O$_3$  \cite{Jani2021}, and AF-AkXs in 
synthetic antiferromagnets \cite{Dohi2019, Legrand2020}.

The appearance of SkXs is not limited to magnetic solids. Indeed, the band structure and electronic topology of moir\'e crystals in 
twisted transition metal dichalcogenides (TMDs) have been understood in terms of a SkX in the layer-pseudospin degree of freedom 
\cite{Wu2019,chen_evidence_2019}. This theoretical proposal has been confirmed by recent scanning tunneling 
measurements \cite{Thompson2024, Liu2024,Zhang2024}. \par

Recent work demonstrates that even centrosymmetric magnetic solids can be attractive platforms for hosting
skyrmions. Nanoscale skyrmions have been reported in centrosymmetric magnets such as bilayer triangular Gd$_2$PdSi$_3$ 
\cite{Kurumaji2019,  Hirschberger2020} and bilayer kagome Gd$_3$Ru$_4$Al$_{12}$ \cite{Chandragiri2016, Hirschberger2019}, 
highlighting the role of magnetic frustration and longer-range Ruderman-Kittel-Kasuya-Yosida (RKKY) interactions in 
creating short-pitch spin spirals and topological textures \cite{Martin2008, Hayami2016,Mitsumoto2022}.
In both materials, the crystal inversion center sits between the Gd-layers, so that inversion symmetry is 
broken in the opposite sense in each layer, but not for the crystal as a whole.

Despite this wide interest in skyrmion crystals and their diverse realizations, the role of electron-electron interactions in 
SkXs is only now starting to be investigated. Recent work shows that magnetic SkXs coupled to dilute electron gases 
can form Chern bands which are quite flat and 
nearly ideal Landau-like levels, potentially leading to the formation of interaction-induced fractional quantum 
Hall states \cite{Paul2023,Reddy2024b}. Similarly, the recent discovery of fractional Chern insulators (FCIs) in the twisted TMDs
\cite{Park2023, Zeng2023, Xu2023} has been understood in terms of electron interaction effects in partially filled SkX (or meron crystal) 
bands \cite{Reddy2024a,Reddy2024b, Morales2024}.

\begin{figure}[t]
 \centering
 \includegraphics[width=0.49\textwidth]{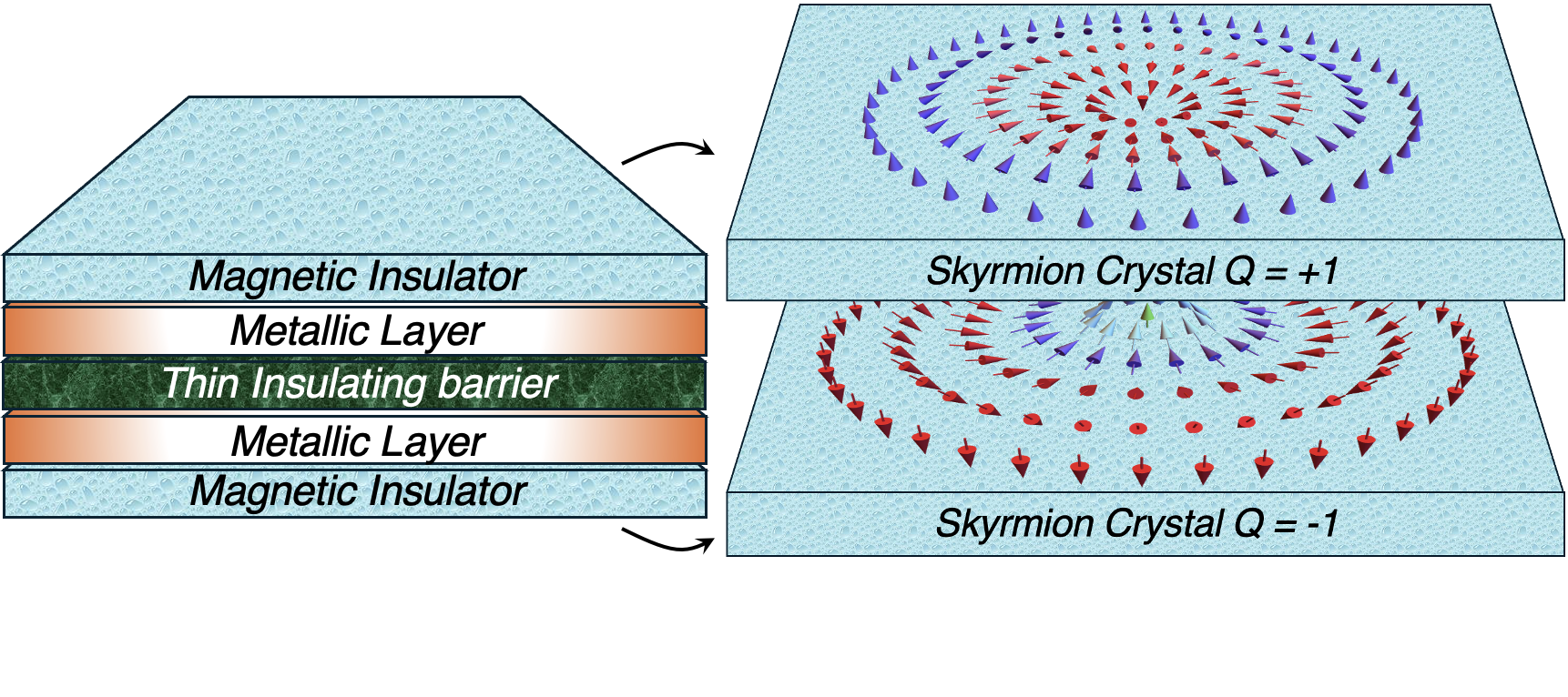}
  \includegraphics[width=0.235\textwidth]{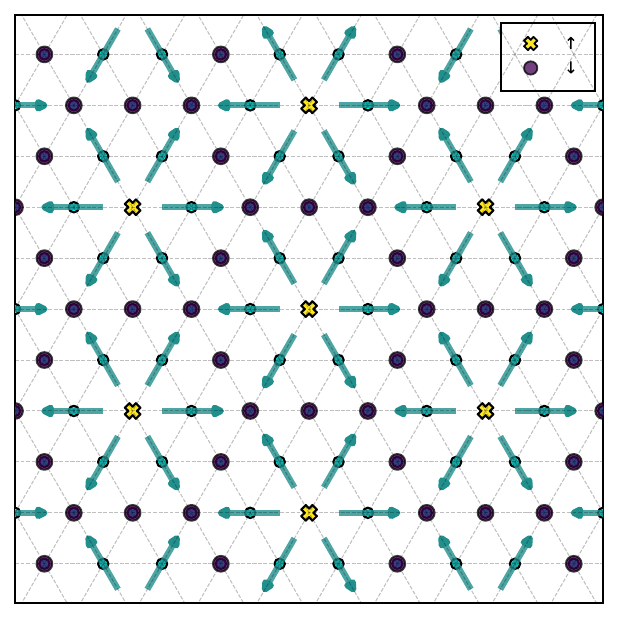}
  \includegraphics[width=0.235\textwidth]{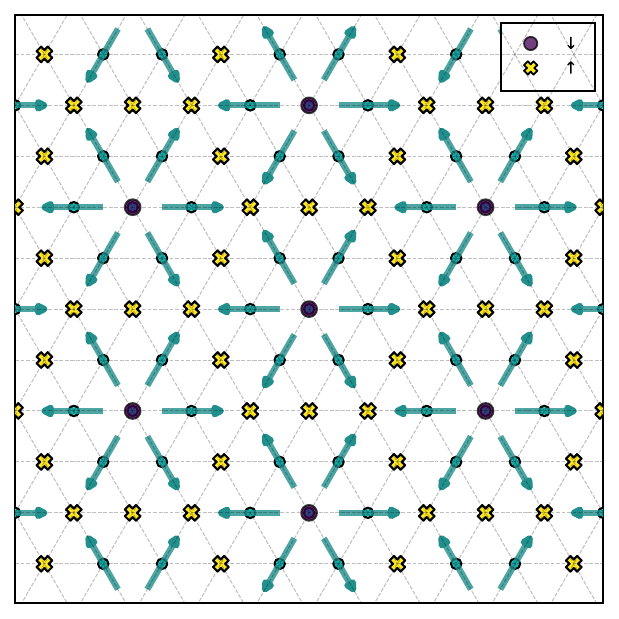}
  \caption{Top: Schematic illustration of a heterostructure with inversion symmetry in the central layer which could lead to interfacial DM interactions of opposite sign at the two metal-magnet interfaces, which may be described by the Hamiltonian in Eq.\ \ref{non-interacting}, with the top and bottom layer N\'eel skyrmions having opposite topological charge. Our key results also remain valid for Bloch skyrmions.
  Bottom: Example of a bilayer Bloch skyrmion texture \(\hat{{\cal S}}_{i, \ell}\) (as defined in Eq.\eqref{weiss}) on the triangular lattice 
  (with 12-site unit cell in our work), and the two 
  layers having opposite ${\cal S}_z$ but the same ${\cal S}_x,{\cal S}_y$.}
\label{fig:layers}
\end{figure}

Motivated by these developments, we study in this Letter the impact of electron-electron interactions in zero field 
bilayer magnetic SkXs. Such bilayer magnetic SkXs with zero net topological charge and a vanishing THE may appear in 
centrosymmetric systems discussed above, or may be 
created in artificial heterostructures as depicted in Fig.~\ref{fig:layers}(a). As a specific example, 
we consider a bilayer SkX on the triangular lattice as depicted in Fig.~\ref{fig:layers}(b),
where $S_z$ is flipped between layer-1 and layer-2 while $S_x,S_y$ remain parallel so that the topological charge 
$\sim \int \bS \cdot \partial_x \bS \times\partial_y \bS$ is opposite
on the two layers. Our focus is on the impact of electron-electron interactions on such SkXs, 
including the impact of both intralayer and interlayer interactions with Hartree-Fock theory. Our main finding is that interaction 
effects can induce various types of broken symmetry states in such SkX states. When the skyrmion bands are somewhat dispersive, 
moderate 
electron interactions lead to an electron crystal which may be viewed as an exciton condensate arising from neighboring Chern bands. 
This crystal does not break the SkX translation symmetry but breaks rotational and
inversion symmetries of the SkX. For flatter Chern bands, even very weak interactions induce 
a weakly layer-polarized phase which also spontaneously breaks inversion symmetry. Both types of broken symmetry phases reveal a
large THE. Our work suggests that it might be interesting to look for such spontaneous inversion symmetry breaking,
perhaps
via optical second harmonic generation accompanying the THE, in the zero field SkX phases of centrosymmetric systems including
artificially engineered heterostructures.

\section{Models of Bilayer skyrmions}

\label{sec:model}

 SkX phases emerge as non-coplanar spin textures due to large magnetic interactions in insulators.    The localized spins have an effective Hund's coupling to the remaining conduction electrons in the following form
\begin{align}
    \mh_0 &=   -t    \sum \limits_{\langle i,j  \rangle, \ell} c^\dagger_{i, \ell, \s} c^\pdg_{j, \ell, \s}  -  J_h    \sum_{i,\ell}
    \hat{{\cal S}}_{i, \ell}\cdot c^\dagger_{i, \ell, \a} \frac{\vec{\s}_{\a \b}}{2} c^\pdg_{i, \ell, \beta} \,,
    \label{non-interacting}
\end{align}  
where \(c_{i, \ell, \s}\) represents an electron at site \(i\) in layer \(\ell\) with spin \(\s\), and all the spin indices are summed over implicitly. The electrons in each layer have a nearest neighbor hopping term, and they are coupled via a local Hund's electron-spin coupling 
to a SkX spin texture which is layer anti-ferromagnetic. The SkX spin configuration $\hat{{\cal S}}_{i, \ell}$ is plotted in Fig.\ref{fig:layers},
where we have chosen ${{\cal S}}^z_{i, 2} = -{{\cal S}}^z_{i, 1}$, i.e. to be flipped between the layers, while 
${{\cal S}}^x_{i, \ell}$ and ${{\cal S}}^y_{i, \ell}$ are chosen to be the same in each layer. Such a configuration ensures that the
topological charge of the skyrmions in each layer is opposite.
Further details of the SkX ansatz are given in Appendix \ref{appendix:classification}; see Eq.\ \ref{weiss}. This Hamiltonian describes an effective Kondo lattice 
between localized spins and conduction electrons. Such an antiferro-SkX states could potentially emerge in magnetic layers of a 
heterostructure stack, as depicted schematically in Fig. \ref{fig:layers}, and realized with micromagnetic simulations \cite{ Zhang2016b} and classical Monte Carlo results \cite{HayamiJpn2023}. The Kondo coupling could then emerge through a proximity effect 
with a metallic layer in the heterostructure stack \cite{Paul2023}. An alternative material realization would be in heavy fermion materials, 
including multipolar SkX configurations, induced by the RKKY interaction coupled to conduction electrons \cite{Mitsumoto2022, zhang2024b}. 
Another material realization would be a layer-SkX state in TMD materials where the moir\'e potential provides the effective Hund's coupling to the electrons \cite{Wu2019}; this connection is further elaborated upon in the Discussion section.
\par

% \begin{figure}[t]
%  \centering
%  \includegraphics[width=0.23\textwidth]{Figures/SkX_Bloch_colored.pdf}
%   \includegraphics[width=0.23\textwidth]{Figures/SkX_Neel_colored.pdf}
%  \caption{ The skyrmion texture chosen on a single layer of triangular lattice with a 12-site unit cell : (a) Bloch skyrmion, and (b) Neel skyrmion. The two textures are related by a simple spin-rotation along \(\hat{z}\) by an angle of \(\pi/2\). The spin-texture on the second layer is symmetric, with the \(z-\)component flipped everywhere.}
% \label{fig:skyrmion}
% \end{figure}

  \begin{figure*}[ht]
     \centering
     \includegraphics[width=0.495\textwidth]{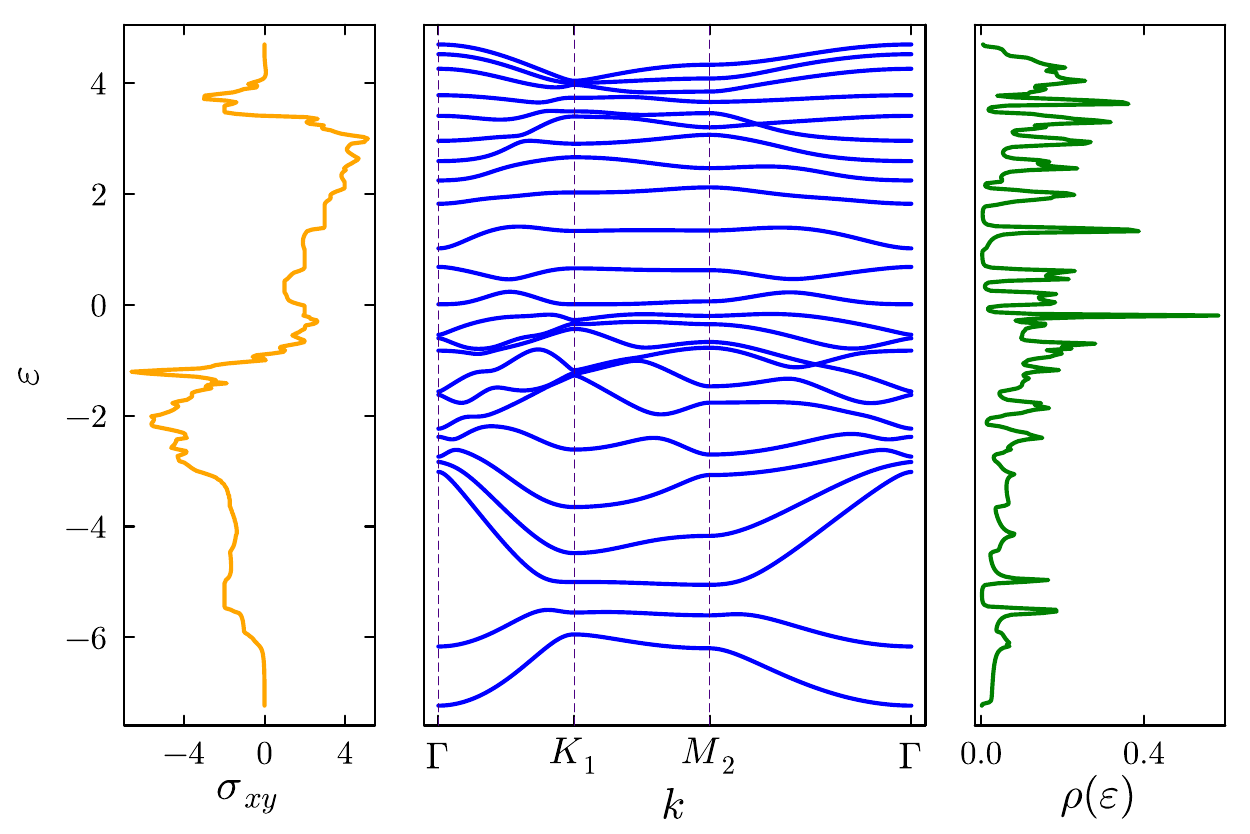}
     \includegraphics[width=0.495\textwidth]{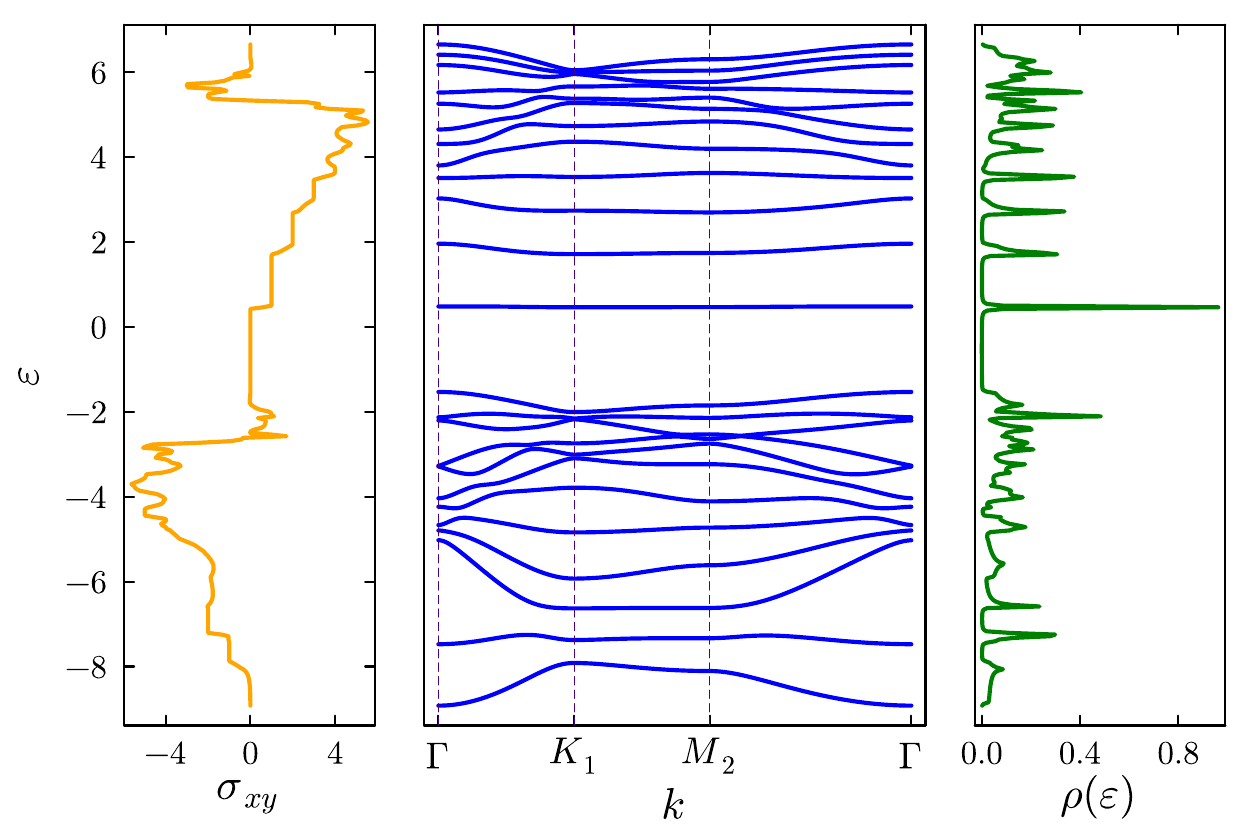}
     \caption{ Band structure of the $N_a = 2$ skyrmion for $J_h/t = 2$ (left) and \(J_h/t=4\) (right). The green plot for each value of the Hund's coupling indicates the density of states $\rho(\varepsilon)$, while the yellow plots show the Hall conductivity $\sigma_{xy}$ (in units of \(e^2/\hbar\)) of the monolayer (with $\sigma_{xy} = 0$ of the non-interacting bilayer throughout). At \(J_h/t=4\), we focus on the Kramers degenerate \(13^{\text{th}}\) band which illustrates a lattice version of a nearly flat, topologically non-trivial (\(C=1\)) band as compared to lower values of the Hund's coupling where its dispersive (see Fig.\ \ref{fig:skyrmion metric measures}).}
    \label{fig:non-interacting}
\end{figure*}

The non-interacting bilayer Hamiltonian has the following symmetries:
\begin{itemize}
    \item Translation : \(T_1\) and \(T_2\) associated with the SkX. In our model, we choose a specific realization of the SkX 
    with a 12-site unit cell in each layer (\(2\sqrt{3}\times2\sqrt{3}\) unit cell); however, the qualitative results that we obtain in this paper
    are not expected to depend on the specific SkX unit cell.
    \item Rotation : The skyrmion texture \(\hat{{\cal S}}_{i, \ell}\) breaks simple lattice rotation symmetry for the electrons. However, there is a residual spin-orbit rotation of the form \(\mathcal{R} = C_{6z}\otimes \mathds{1}_{\ell}\otimes R_{\hat{z}}(\pi/3)\), where \(C_{6z}\) is a lattice rotation by an angle of \(\pi/3\) around an axis along \(\hat{z}\) passing through the center of the skyrmion. \(\mathds{1}_{\ell}\) acts as identity on the layer index, while \(R_{\hat{z}}(\pi/3)\) is a spin-rotation of the electron around \(\hat{z}\).
    \item Centrosymmetry : While the bilayer lattice has a inversion symmetry around a central point in between the layers \(\mathcal{I}\), the skyrmion texture \(\hat{{\cal S}}_{i, \ell}\) again breaks this symmetry, along with time reversal, \(\mathcal{T}\). However, the composite symmetry \(\mathcal{C} = \mathcal{T}\cdot\mathcal{I}\) is still preserved. Furthermore, the inversion symmetry \(\mi\) itself can be decoupled into \(\mi=C_{2z}\otimes\mm_{xy}\), where \(C_{2z}\) is a lattice rotation by an angle of \(\pi\) around an axis along \(\hat{z}\) passing through the center of the skyrmion, while \(\mm_{xy}\) is a {\it lattice} mirror symmetry \textit{w.r.t} the mirror \(xy\) plane which lies between the layers but does not act on spin degrees of freedom.
    \item In-plane Mirror : The skyrmion texture has a composite spin-orbit coupled version of mirror symmetry \(\mathcal{M}_{xz}\) and \(\mathcal{M}_{yz}\). 

\end{itemize}

As demonstrated in Fig.\ \ref{fig:non-interacting}, the SkX fragments the itinerant electron bands into topological minibands \cite{Reddy2024b}. 
The spin texture also instills a non-trivial Berry curvature onto the minibands. 
with Chern number $C \neq 0$ generically. However, due to the centrosymmetry, \(\mathcal{C}\), present in the full bilayer Hamiltonian, each band is doubly degenerate with a band of opposite layer, with opposite Chern numbers $C_{n}^{\ell} = - C_{n}^{\bar{\ell}}$ as well. 
 The topology of the resulting degenerate two-band subspace resembles a generalized Kane-Mele model \cite{Kane2005} which leads to a topological spin Hall effect (TSHE) \cite{Gobel2017a}. We now turn to the observable consequences of this topology. \par

\section{Topology of skyrmion bands}

\subsection{Topological Hall Effect}
The Kubo formula gives the conductivity $ \sigma_{ij}(\mu)$ at a given chemical potential \(\mu\) as
\begin{equation}
\sigma_{ij}(\mu)=\sum_{\bk}\sum_{\e_{\a}\leq\mu} \sum_{\e_{\b}>\mu}\frac{\left\langle\psi_\alpha\left|\partial_i \mh(\bk)\right| \psi_\beta\right\rangle\left\langle\psi_\beta\left|\partial_j \mh(\bk)\right| \psi_\alpha\right\rangle}{\left(\e_{\alpha}(\bk)-\e_{\beta}(\bk)\right)^2}.
\end{equation}
Even in a partially filled band (metallic state), this can be rewritten as 
\begin{equation}
  \sigma_{x y}\left(\e_{\mathrm{F}}\right)=\frac{e^2}{\hbar} \sum_n \dk \Omega^{n}(\bk) n_F\left(\e_{n}( \bk)-\e_F\right), 
\end{equation}
where the Berry curvature is the gauge invariant quantity given by \(\W^{n}(\bk) = \partial_x \ma^{n}_{y}(\bk) - \partial_y \ma^{n}_{x}(\bk)\) from the Berry connection \(\ma^{n}_{i}(\bk)\)\cite{haldane2004}. For fully filled isolated bands, the Hall conductivity matches the Chern number as $2 \pi \sigma_{xy} =  C$.  The Hall response of individual layers of the AF-SkX cancels with one another due to the centrosymmetry \(\mathcal{C}\), leaving a net-zero THE. The Hall response of an individual layer has been plotted in Fig.\ref{fig:non-interacting}.

\subsection{Quantum Geometric Tensor of Skyrmion Bands}
\begin{figure*}[!ht]
 \centering
 \includegraphics[width=0.99\textwidth]{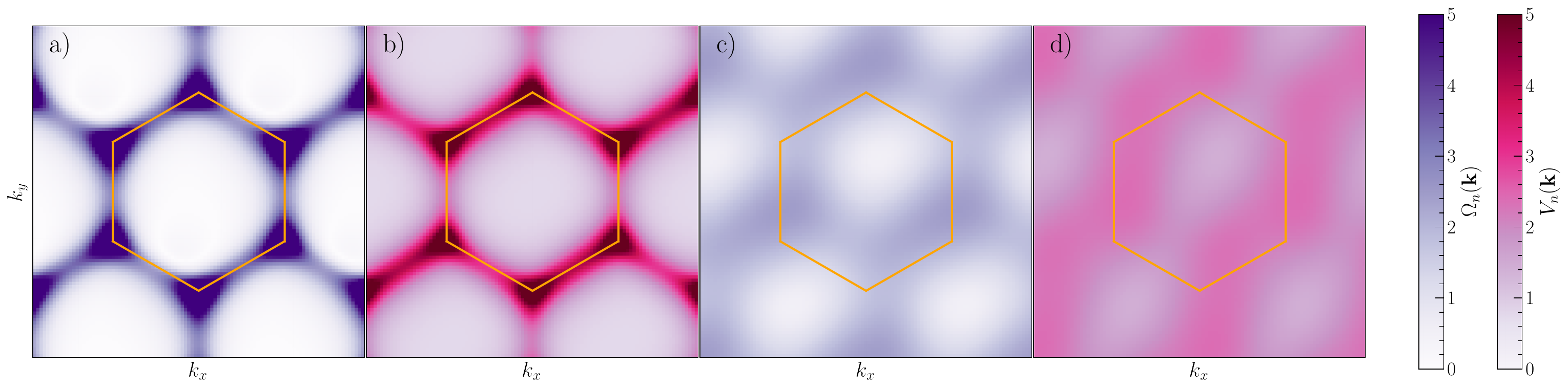}

 \caption{ (a) The momentum resolved  Berry curvature $\W^n(\bk)$, and (b) the quantum volume $V^n(\bk)$ for the first SkX band \(n=1\) at \(J_h/t=4\) which is dispersive. There is clear structure as seen by the peaks at the high-symmetry \(\bk=K,K'\) points. (c) The momentum resolved  Berry curvature $\W^n(\bk)$, and (d) the quantum volume $V^n(\bk)$ for the \(13^{\text{th}}\) SkX band at \(J_h/t=4\) which is a topological flat band, and is comparatively featureless and uniform. Both the bands have the same Chern number, \(|C|=1\), and hence every measure is drawn on the same scale. 
 The yellow hexagon marks the SkX Brillouin zone. For comparison, the same plots at a lower Hunds coupling of \(J_h/t=2\) are plotted in Fig.\ \ref{fig:skyrmion metric Jh=2} in Appendix.\ \ref{appendix:J_2}.}
\label{fig:skyrmion metric}
\end{figure*}

More general topological responses of the bands can be extracted through geometric properties of the crystalline wavefunctions encapsulated in the quantum geometric tensor (QGT) \cite{Roy2014}.
The QGT can be decomposed as follows into symmetric and anti-symmetric components.
\begin{equation}
T^{n}_{\a \b}(\bk) \equiv g^{n}_{\a \b}(\bk)-\frac{i}{2} \Omega^{n}_{\a \b}(\bk),
\end{equation}
where \(g^{n}_{\a\b}(\bk)\) is the quantum metric of an isolated band \(n\) evaluated at momenta \(\bk\), and \(\a,\b\in \{x, y\}\). This defines a Fubini-Study metric on the projective space of wavefunctions \cite{Wang2021}.  The imaginary component $ \Omega^{n}_{\a \b}(\bk)$ is the Berry curvature tensor whose off-diagonal element is the usual Berry curvature \(\W^{n}(\bk)\), responsible for the quantum Hall response.  
The quantum metric captures the propensity for Chern bands to form Landau-Level analogs, quantified by the trace condition \cite{Roy2014}
\begin{equation}
    \label{Trace condition}
    \frac{1}{2}\Tr\left(g^n(\bk)\right)\geq\sqrt{\det(g^n(\bk))} \geq \Omega^n(\bk)\,,
\end{equation}
where \(W^n(\bk)\equiv\Tr\left(g^n(\bk)\right)\) is defined as the quantum weight of band \(n\) at momentum \(\bk\), while \(V^{n}(\bk)\equiv\sqrt{\det(g^n(\bk))}\) is called the quantum volume. \par

The second bound in Eq.\eqref{Trace condition} is saturated for two-band Hamiltonians (such as the Haldane model), while both the bounds are saturated for isotropic ideal states, such as the lowest Landau level (LLL) or combined $n$-Landau levels. 
For Landau level analogs, the quantum metric plays a significant role in determining the localization and stability of the electronic states.
A flat band with a uniform quantum metric and non-zero Berry curvature can mimic the properties of Landau levels, leading to lattice analogs of fractional quantum Hall states, or fractional Chern insulator states \cite{Regnault2011, Roy2014, Wang2021}. 
A completely uniform quantum metric $\nabla g(\bk) = 0$ implies that the density matrices of these wavefunctions obey a generalization of the $W_\infty$ algebra \cite{Parameswaran2012,  Roy2014}.
This property should even hold  if the trace condition is not satisfied and the bands are more analogous to higher Landau levels \cite{Roy2014}.\par
As shown in Fig. \ref{fig:non-interacting}, and in agreement with previous work that studied the topology of SkX bands\cite{Paul2023}, we find a nearly perfectly flat band (where flatness can be quantified as the ratio of bandwidth to the band gap), the $13^\text{th}$ band, just above half-filling at \(J_h/t=4\). We demonstrate the momentum resolved findings in Fig. \ref{fig:skyrmion metric} for this band with a flatness of \(\approx 0.01\) and \(C=1\). For comparison with a dispersive band, we also plot the same quantities for the lowest band (\(n=1\)) which has a flatness of \(1.86\) and $C = 1$. We find the Berry curvature  $\W^n(\bk)$, and quantum volume $V^n(\bk)$ are peaked at high-symmetry $\mathbf{K}$ points. For the flat band $\W^n(\bk)$, and  $V^n(\bk)$ are nearly uniform,  although they maintain a constant proportionality. We see a variation of the Berry curvature  throughout the Brillioun Zone of \(\s(\Omega) \approx 0.5 \).  This uniformity suggests that the flat band is an excellent candidate for realizing Landau level analogs. \par

\begin{figure}[t]
 \centering
 \includegraphics[width=0.45\textwidth]{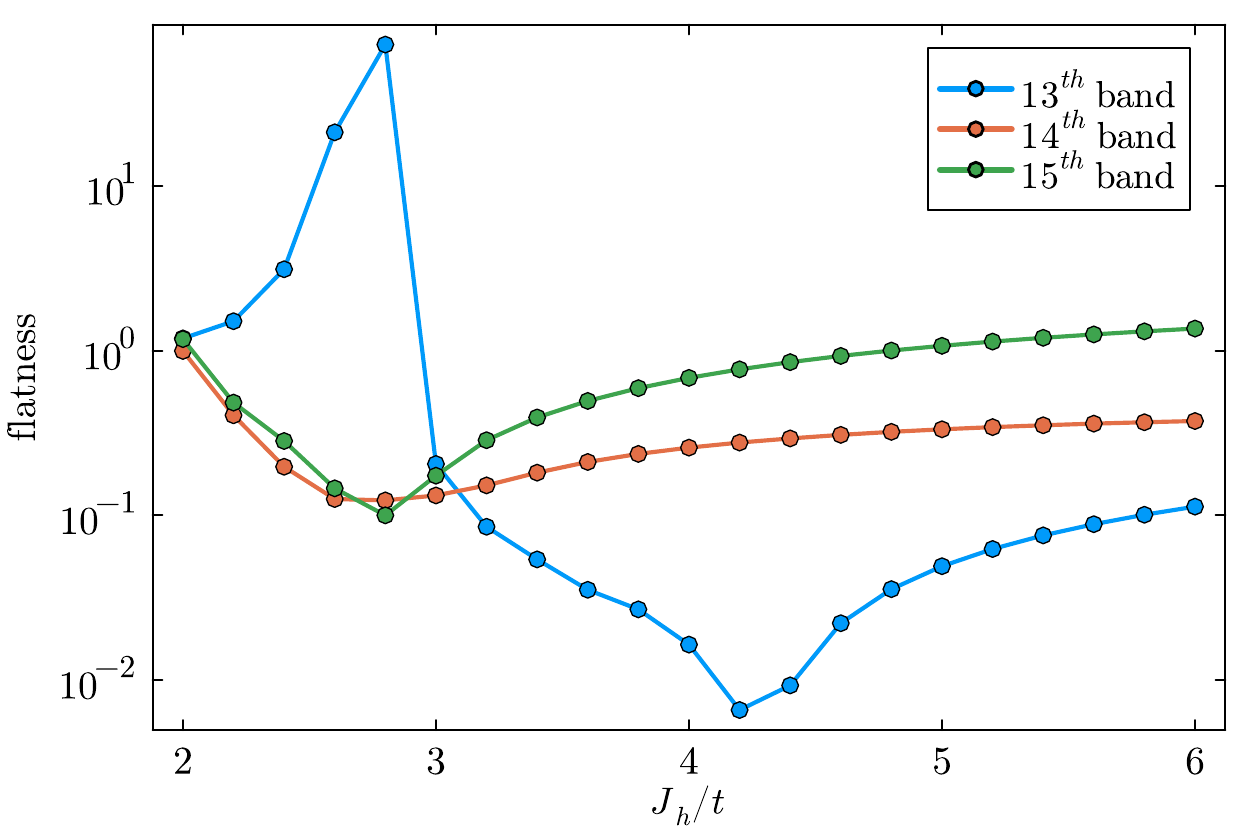}
 \includegraphics[width=0.45\textwidth]{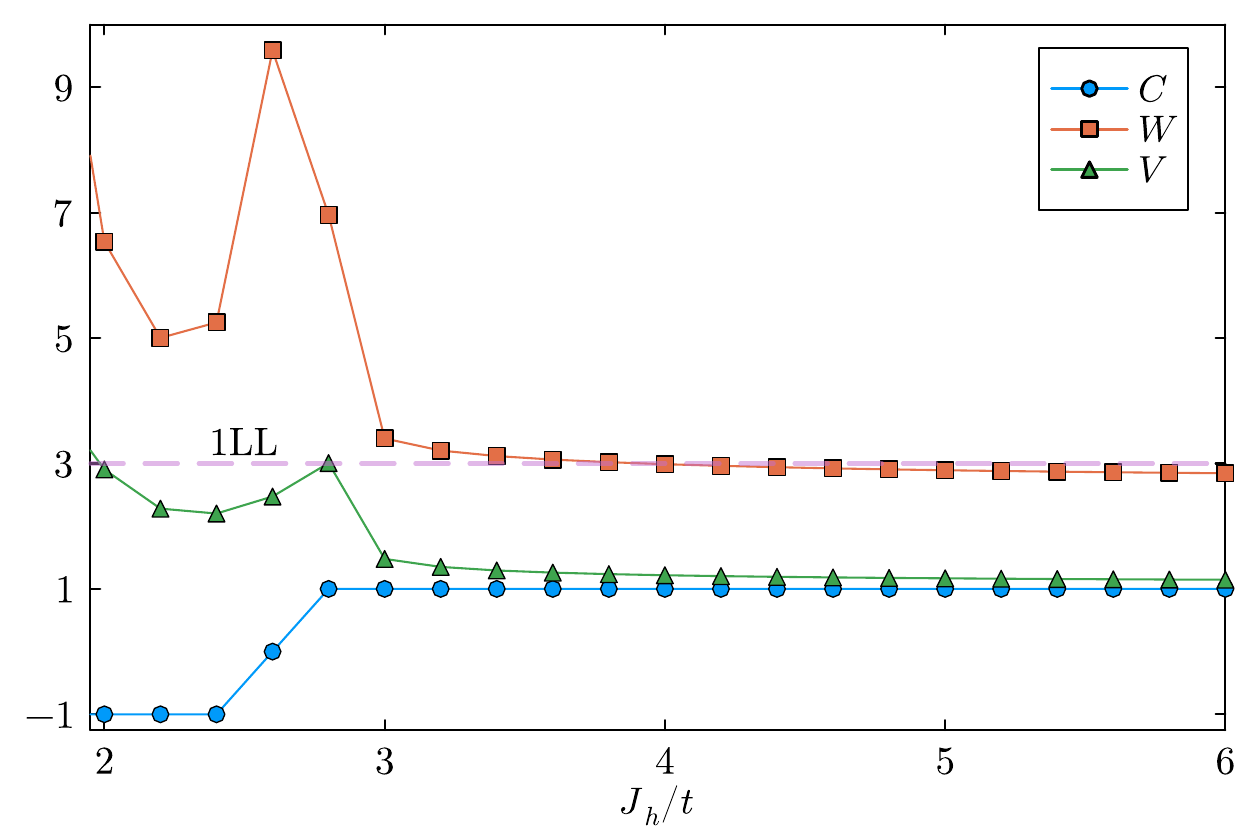}
 \caption{ Various properties of the skyrmion flat bands as the Hunds coupling \(J_h\) is tuned : (a) The flatness (defined as the ratio of bandwidth to the minimum gap to a band below or above) of the \(13^{\text{th}}\), \(14^{\text{th}}\), and \(15^{\text{th}}\) bands in the monolayer model. As can be seen, there is a minima of the flatness of the \(13^{th}\) band near \(J_h/t\approx 4.2\), and the flatness is \(\lessapprox 0.01\) for a range of values \(J_h/t\in (4, 4.5)\). The jumps seen are due to band touchings as we tune the Hund's coupling. (b) Some quantum metric measures of the \(13^{\text{th}}\) band. \(C\) refers to the Chern number, \(W\) refers to the quantum weight, and \(V\) refers to the quantum volume. For reference, the first Landau level would have a quantum weight of 3.}
\label{fig:skyrmion metric measures}
\end{figure}

One can also define the total quantum weight and total quantum volume as the integral over the entire Brillouin zone of the respective quantities, $W_n = \frac{1}{2\pi}\dk W_n(\bk) $ and $V_n = \frac{1}{2\pi}\dk V_n(\bk) $, similar to how the Chern number is the integral of the Berry curvature. These quantum properties of the \(13^{th}\) band and nearby flat bands as one tunes $J_h$ are demonstrated in Fig.\ \ref{fig:skyrmion metric measures}. For the $13^\text{th}$ band at \(J_h/t=4\),  we find $W_n = 2.99$, near the ideal value of $3$ for the first Landau Level and in close parallel with similar work on Landau level analogs with lower lying states in a SkX \cite{Morales2024, Reddy2024b}. We find $V_n  = 1.22$, in some agreement of a 1LL value of $ 1.5$. Lastly, in the same figure, we find that the flatness of the bands is non-monotonic with respect to $J_h$. For the rest of the discussion in this paper, we will focus on $J_h/t = 4$ near the optimum region, as well as \(J_h/t=2\) for comparison with a dispersive band.

\section{Interaction induced electronic instabilities}
We now turn to the role that the skyrmion crystal and electronic interactions have on charge ordering and band topology.
To understand possible charge orders, we define the layer-resolved charge density in momentum space as
 \begin{equation}
 \label{rho(q)}
     n_{\ell}(\bQ) = \sum_{\s}\dr\: e^{i \bQ\cdot\br} \expval{c^{\dagger}_{\br, \ell, \s}c^\pdg_{\br, \ell, \s}}\,.
 \end{equation}
 Using this, we can define the layer-combined charge and polarization at a given momentum as \(n(\bQ) = n_{1}(\bQ)+n_{2}(\bQ)\), and \(\Delta n(\bQ) = n_1(\bQ)-n_2(\bQ)\). Note that \(n(\bQ=\G) = 2\bar{n}\) represents the combined uniform charge density per site, while \(\Delta n(\bQ=\G)\) represents the uniform polarization per site, where \(\G = (0, 0)\). To study the non-uniform part of the charge distribution, we also plot the Fourier transform of the charge modulation,  \(n_{\ell}(\bQ)-\overline{n}_{\ell}\). These peaks would indicate additional charge ordering or modulated polarization. \par

\begin{figure}[!ht]
 \centering
 \includegraphics[width=0.495\textwidth]{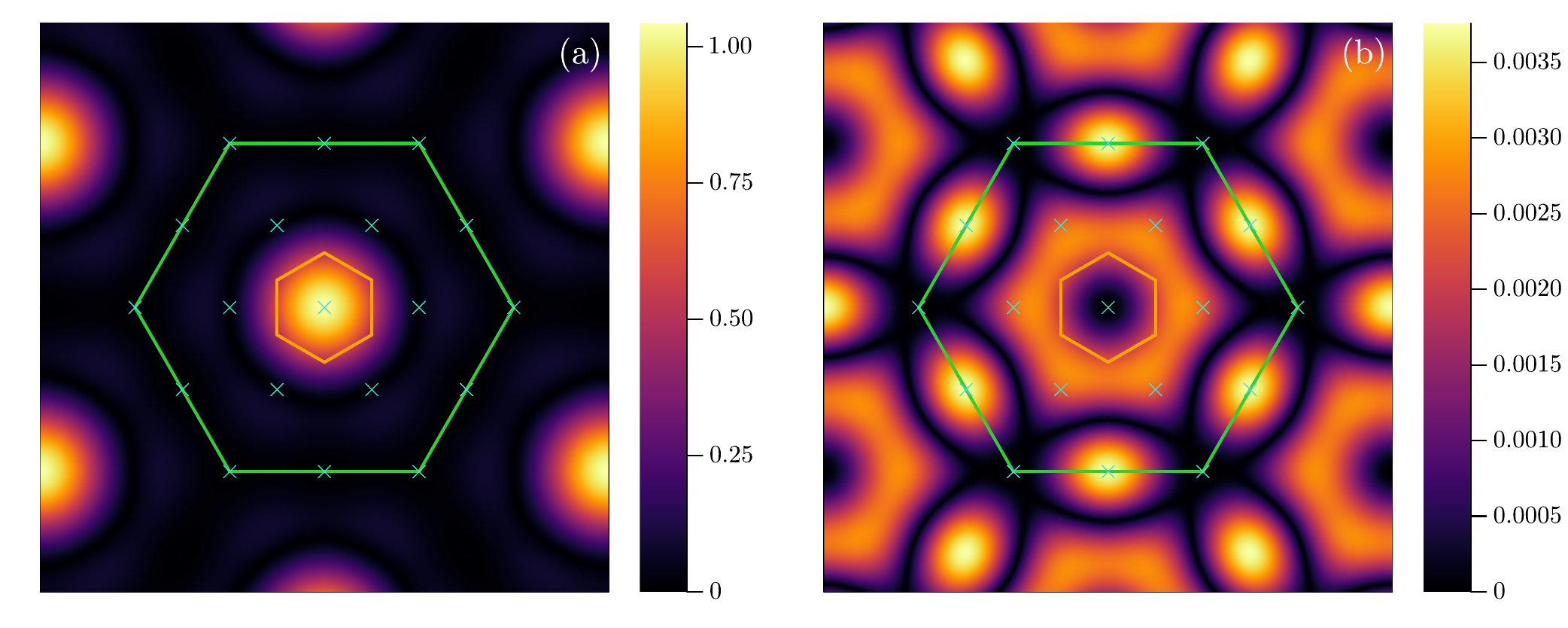}
 \caption{ The charge distribution in momentum space in the presence of the skyrmion at \(J_h/t=4\) when the \(13^{\text{th}}\) degenerate band is half-filled at \(\bar{n}= 25/48\). (a) The absolute value of the total \(|n(\bQ)|\) in the triangular Brillouin zone. (b) The finite $q$ piece of part (a).  The green hexagon marks the triangular Brillouin zone, while the smaller yellow one is that of the SkX unit cell. The \(\times\) points mark the reciprocal lattice sites corresponding to the SkX unit cell.}
\label{fig:skyrmion charges}
\end{figure}

The bare SkX state provides a small amount of charge redistribution, which is documented in Fig.\ \ref{fig:skyrmion charges}. 
It is clear from the peak at \(\bQ=0\) that most of the charge is distributed uniformly. 
However, from the modulated charge distribution, we see there are small but non-zero charge distribution at other momenta like the \(M\) points, as well as the SkX ordering vector, which are the \(K/2\) points. 

\subsection{Electron Repulsion}
We consider the next two leading contributions parameterized by short-range Hubbard-type interactions
\begin{equation}
\begin{gathered}
    \mh_{int} = U \sum_{i}\sum_{ \ell} n_{i, \ell} n_{i , \bar{\ell}} + V \sum_{\langle i, j\rangle}\sum_{ \ell} n_{i, \ell} n_{j, \ell} .
    \end{gathered}
\end{equation}
These are repulsive nearest-neighbor density-density interactions, parameterized by $V$ and the inter-layer density-density repulsion, parameterized by $U$.
In our mean-field analysis, we will not consider the electronic feedback on the SkX background spins or potential magnetic orderings. We instead consider potential charge instabilities of the itinerant electrons.  

We solve the full Hamiltonian $\mh_0 + \mh_{\text{int}}$ with a generalized mean-field decomposition in the Hartree-Fock channels. In particular, we decompose into bond-dependent intra and inter-layer hopping, as well as a density term for each site separately. The resulting mean-field Hamiltonian decomposition is $\mh_{MFT} = \mh_0 + \mh_U + \mh_V$, with
\begin{eqnarray}
       \mh_U    &=&   U \sum_{i,\ell}\left( \bar{n}^\pdg_{i, \ell}c^{\dagger}_{i, \bar{\ell}, \s}
    c^\pdg_{i, \bar{\ell}, \s}  -  \c_{ii}^{\ell\bar{\ell}}c^{\dagger}_{i, \bar{\ell}, \s}c^\pdg_{i, \ell, \s}\right)\\
      \mh_V    &=&    V  \sum_{\langle i, j\rangle,\ell}\left(\bar{n}^\pdg_{i, \ell}c^{\dagger}_{j, \ell, \s}
   c^\pdg_{j, \ell, \s}  - \c_{ij}^{\ell\ell}c^{\dagger}_{j, \ell, \s}c^\pdg_{i, \ell, \s}  +  i\leftrightarrow j\right)\,,
\end{eqnarray} where
where \(\c_{ij}^{\ell_1\ell_2} = \expval{c^{\dagger}_{i, \ell_1, \s}c_{j, \ell_2, \s}}\). The Hartree terms give rise to site-dependent and layer-dependent chemical potentials given by the expectation values of the densities \(\bar{n}_{i, \ell, \s}\), while the Fock terms give rise to inter-layer hopping \(\c_{ii}^{\ell\bar{\ell}}\), and renormalized intra-layer hopping \(\c_{ij}^{\ell\ell}\).\par
 
\subsection{Layer Polarized Phase}
 This time-reversal invariant system is protected by a combined centrosymmetry $\mathcal{C} = \mathcal{T}\cdot\mathcal{I}$. 
If the centrosymmetry \(\mathcal{C}\) is spontaneously broken, then the system would reveal a `hidden' topological invariant.  
 The simplest symmetry broken state with finite THE would be a layer polarized state where one of the SkX states is preferred. This would be a state which spontaneously breaks \(\mm_{xy}\), and therefore breaks the centrosymmetry, \(\mc\), which relates the two degenerate bands. A similar phenomenon occurs for valley polarization in moir\'{e} materials with a spontaneous anomalous Hall effect \cite{Zhang2019,Repellin2020, Bultinck2020, Bultnick2020b,  XieY2022, Crepel2023}. In such cases, the composite symmetry is instead $ \mt\cdot C_{2z} $.

 \begin{figure}[t]
     \centering
     \includegraphics[width=0.49\textwidth]{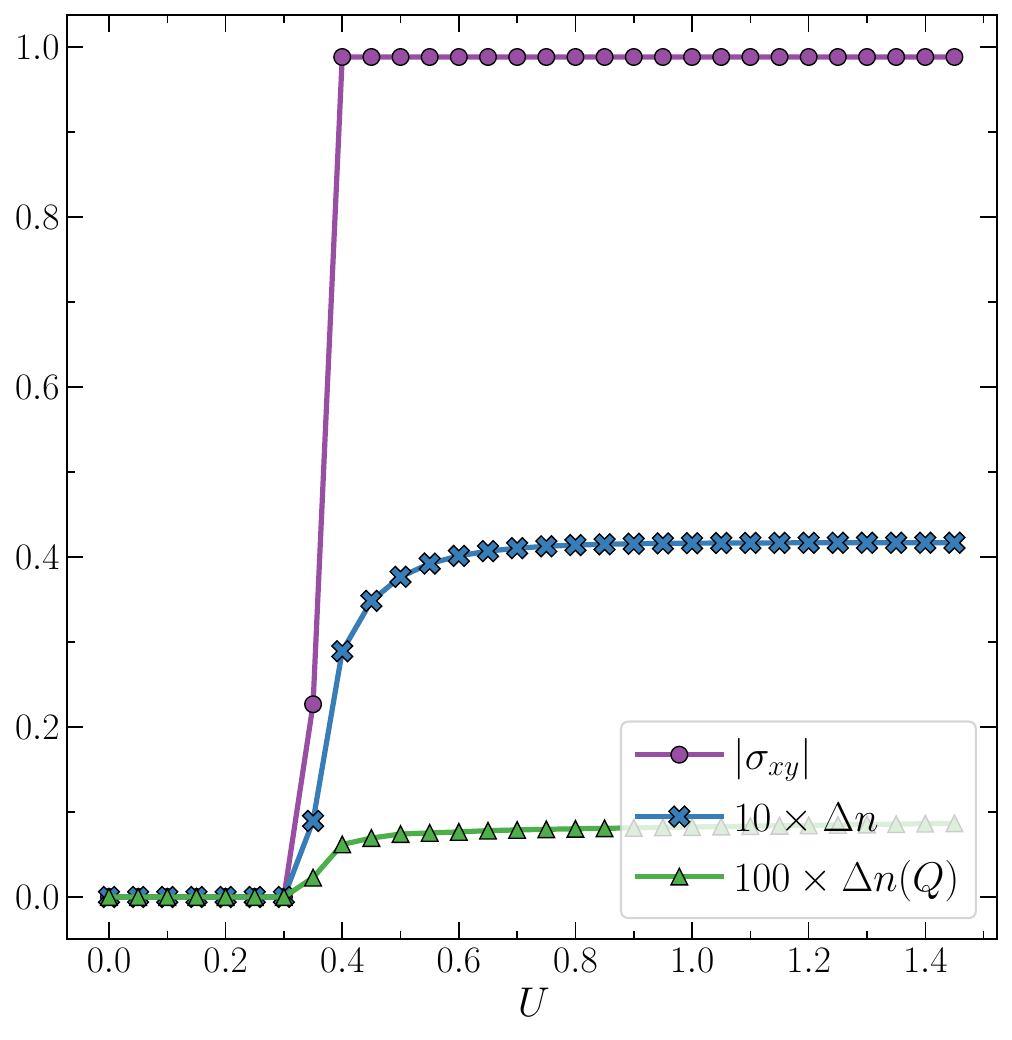}
     \caption{For  $J_h/t = 4$, The topological Hall response $\sigma_{xy}$, the polarization $\Delta n$, and the charge order $\Delta n (Q)$  that emerge as a continuous transition with increasing the interlayer repulsion, $U$. Here we consider $V = 0$. The underlying skyrmion ordering vector is $\bQ = \bK/2$. }
    \label{one-band}
\end{figure}
 For the rest of our discussion, we focus on the specific filling $\bar{n} = 25/48 \approx 0.521$. This corresponds to half-filling of the combined Kramers-degenerate $13^{\text{th}}$ band of the bilayer. 
 For the polarized case, this means the lower band will be fully occupied, driving the system to a genuine $\mathbb{Z}$ topological insulator with a quantized topological Hall response $2 \pi  \sigma_{xy} =  C$. \par
 Fig.\ \ref{one-band} demonstrates a net layer polarization due to splitting of the degenerate bands.  Subtracting this uniform piece demonstrates residual charge distribution at incommensurate $\bQ$, proportional to the SkX as shown in Fig.\ \ref{fig:skyrmion charges_J_4_U_2} in Appendix.\ \ref{appendix:structure factor}. 
 As such, one band becomes fully filled and the system exhibits a quantized Hall response, a spontaneous topological Hall effect. 
 The polarization order occurs at $\bQ = \Gamma$, with uniform polarization in one layer. \par
\begin{figure}[t]
     \centering
     \includegraphics[width=0.5\textwidth]{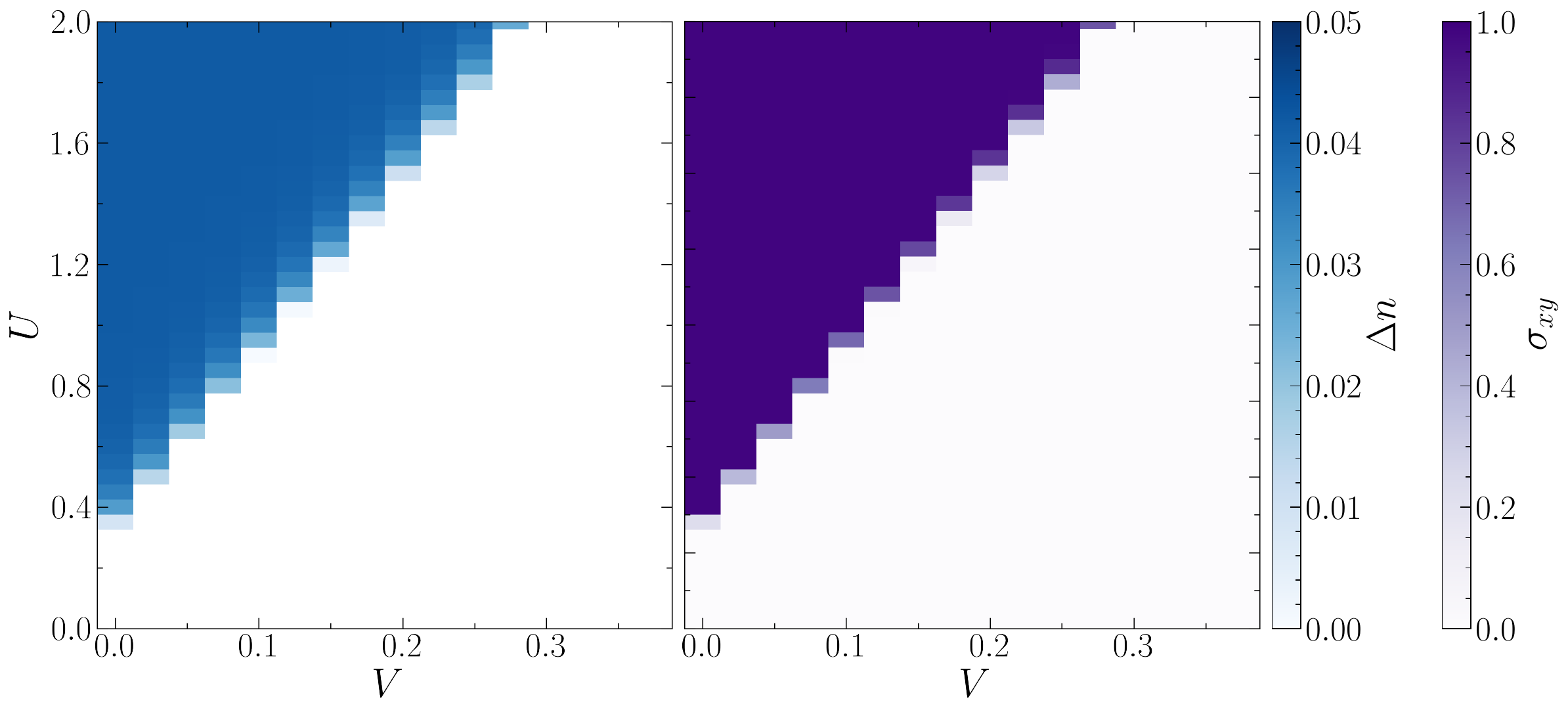}
     \caption{The polarization $\Delta n$ and topological Hall response $\sigma_{xy}$ as a function of $U$ and $V$ for the flat-band under discussion for $J_h = 4$.}
    \label{sigmaxy}
\end{figure}

 The effect of nearest-neighbor $V$ is demonstrated in Fig.\ \ref{sigmaxy}.  $V$ suppresses the polarization due to $U$. This is due to the intra-layer repulsion favoring order at $\bQ \neq \G$. 

 This is in contrast to Landau levels with instabilities towards charge density waves \cite{Koulakov1996, Moessner1996, Rezayi1999}. It is also in contrast to results on SkX-induced Chern bands with projected long-range Coulomb interactions \cite{Reddy2023} for fillings per band $\nu \leq \frac{1}{2}$. For short-range interactions, we find no such instabilities away from half-filling. The primary effect of tuning to other fractional fillings (both above and below $\nu = \frac{1}{2}$) is to increase the critical $U$ for the polarization transition. This is due to the fact that the projected Coulomb interaction, especially into higher Landau levels, has a considerable amount of structure as compared to the short-range interactions considered here \cite{Aleiner1995}.   This is confirmed through our RPA calculations in Appendix \ref{appendix:susceptibilities}.  The absence of order at the mean-field level leaves open the possibility of topological order induced by interactions \cite{Regnault2011}.  The case for dispersive bands at smaller $J_h$ has dramatic differences.

  \begin{figure}[t]
     \centering
     \includegraphics[width=0.475\textwidth]{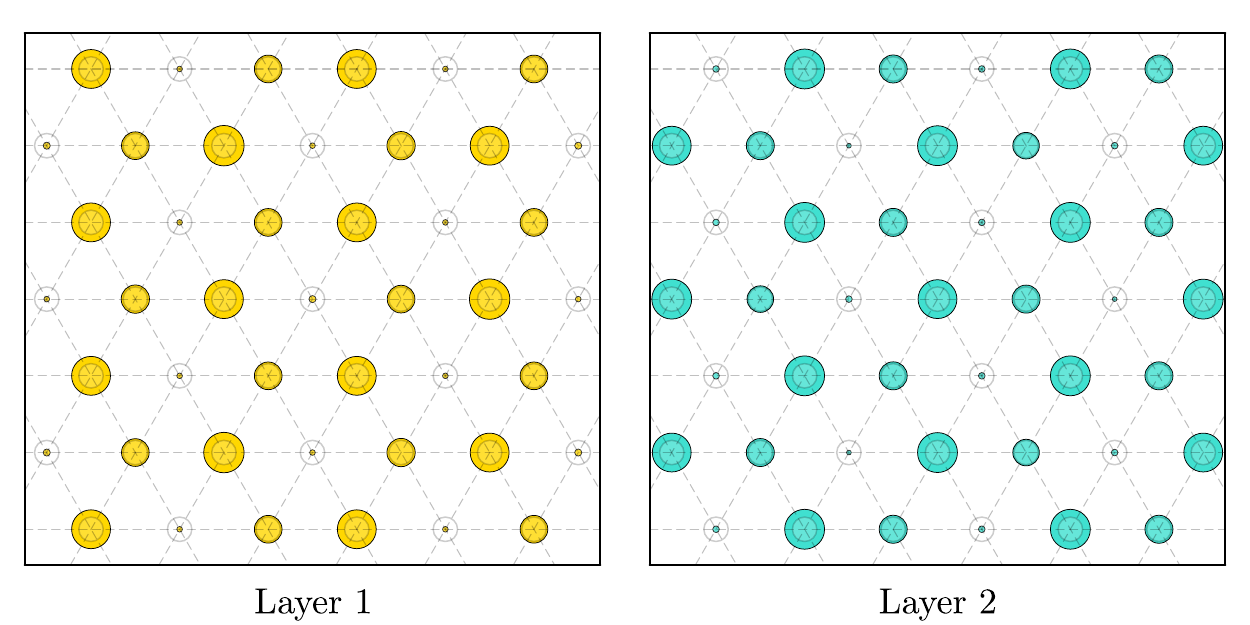}
     \caption{Real space charge order for each layers at $J_h/t = 2$, \(V/t=1.3\), and \(U/t=0.6\) at filling \(\bar{n} = 25/48\). Each site shows the mean-field expectation value of total charge density (both spins) with the radius of the circles being proportional to the actual values. A translucent black circle is plotted on each site showing the uniform total density for reference.  }
    \label{fig:charge_configuration}
\end{figure}

\subsection{Charge Ordered Phase}

\begin{figure}[t]
     \centering
     \includegraphics[width=0.45\textwidth]{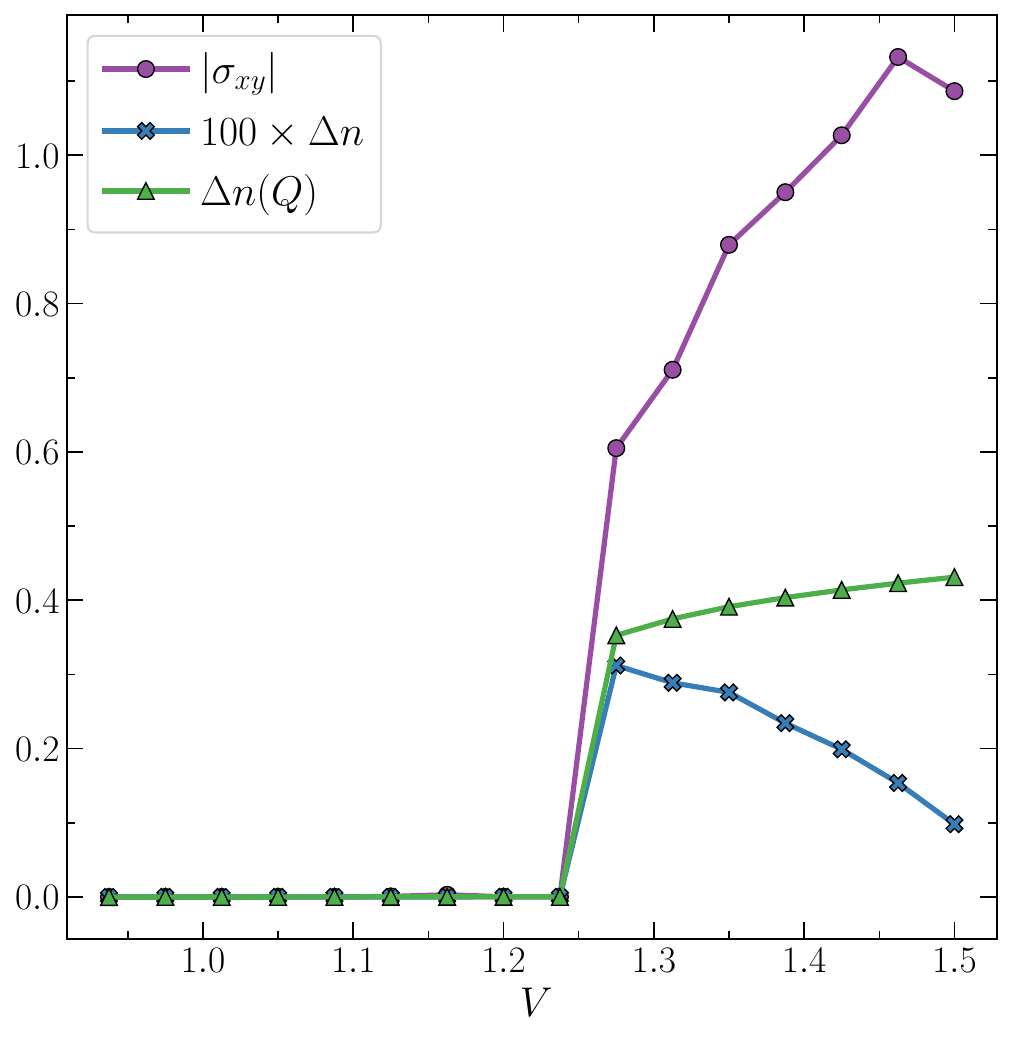}
     \caption{For  $J_h/t = 2$, The topological Hall response $\sigma_{xy}$, the polarization $\Delta n$, and the charge order $\Delta n (\bQ)$ that emerge as a continuous transition with increasing the intralayer repulsion, $V$. the ordering vector is $\bQ = \bK$. The data is plotted for $U/t = 0.6$ as a representative sample. }
    \label{rotation}
\end{figure}

At $J_h/t = 2$, the $13^\text{th}$ band is not isolated and the system is metallic. 
For half-filling this band ($\bar{n} = 25/48$),  the system is stable to layer polarization for small $U$, $ (U \lesssim 2)$. 
However, for moderate $V/t \sim 1.5$, there is a rotational symmetry breaking transition ($C_{6z} \rightarrow C_{3z}$), demonstrated in Fig.\ \ref{fig:charge_configuration} and  Fig.\ \ref{rotation}.  Here instead it is the breaking of \(C_{2z}\)  which breaks the centrosymmetry \(\mc\) and leads to a spontaneous topological  Hall effect. The charge modulation in Fig.\ \ref{fig:skyrmion charges_J_2_V_1.5} in Appendix \ref{appendix:structure factor} demonstrates this rotational symmetry breaking. 
Throughout the ordered phase, there is a considerable unquantized Hall response $\s_{xy}$.
The large number of bands intersecting prevents any quantitative conclusions to be drawn from precise values of $\s_{xy}$. 
This transition occurs due to band-touching and reformation points rather than a simple degeneracy lifting that occurs for larger $J_h$. This phenomenology resembles Landau level mixing.  At large $J_h$, the band gaps are too large and the Fermi surface is too featureless for any weak-coupling instabilities to crystalline symmetry broken states.   \par
For the dispersive case ($J_h/t = 2$), each layer forms an emergent approximate honeycomb lattice with distinct A,B sublattices. At moderate $U$, when overlayed with one another, these form two triangle lattices with an increased unit-vector. The layer-resolved density is shown in Fig.\ \ref{fig:charge_configuration}. 
The overall system has a small net polarization $\Delta n$. The primary ordering is the intralayer charge reconstruction.
% The resulting state however is different from other $\sqrt{3} \times \sqrt{3}$ ordering, such as those in moir\'e materials, in that the ordering here requires mixing of higher bands rather than inter-band coherence \cite{Bultinck2020, Bultnick2020b}.  As the band-gap of Chern bands in TMDs is $\mathcal{O}(\Delta) < 10 $meV, the importance of interaction effects for band-mixing should not be neglected. 
% Future work will determine alternative orders could in fact forms excitons and therefore IVC states with $\sqrt{3} \times \sqrt{3}$ ordering.  
The consequences of charge ordering are exhibited in Fig.\ \ref{rotation}.  
The Hall response $\sigma_{xy}$ that emerges after spontaneous symmetry breaking is unquantized and exists due to partial-filling of the reconstructed bands.  
We have also explored the system at various integer-fillings. Such broken-symmetry states exist at strong-coupling, although the fully-filled bands form trivial crystals.
 \par
\begin{figure}[t]
     \centering
     \includegraphics[width=0.5\textwidth]{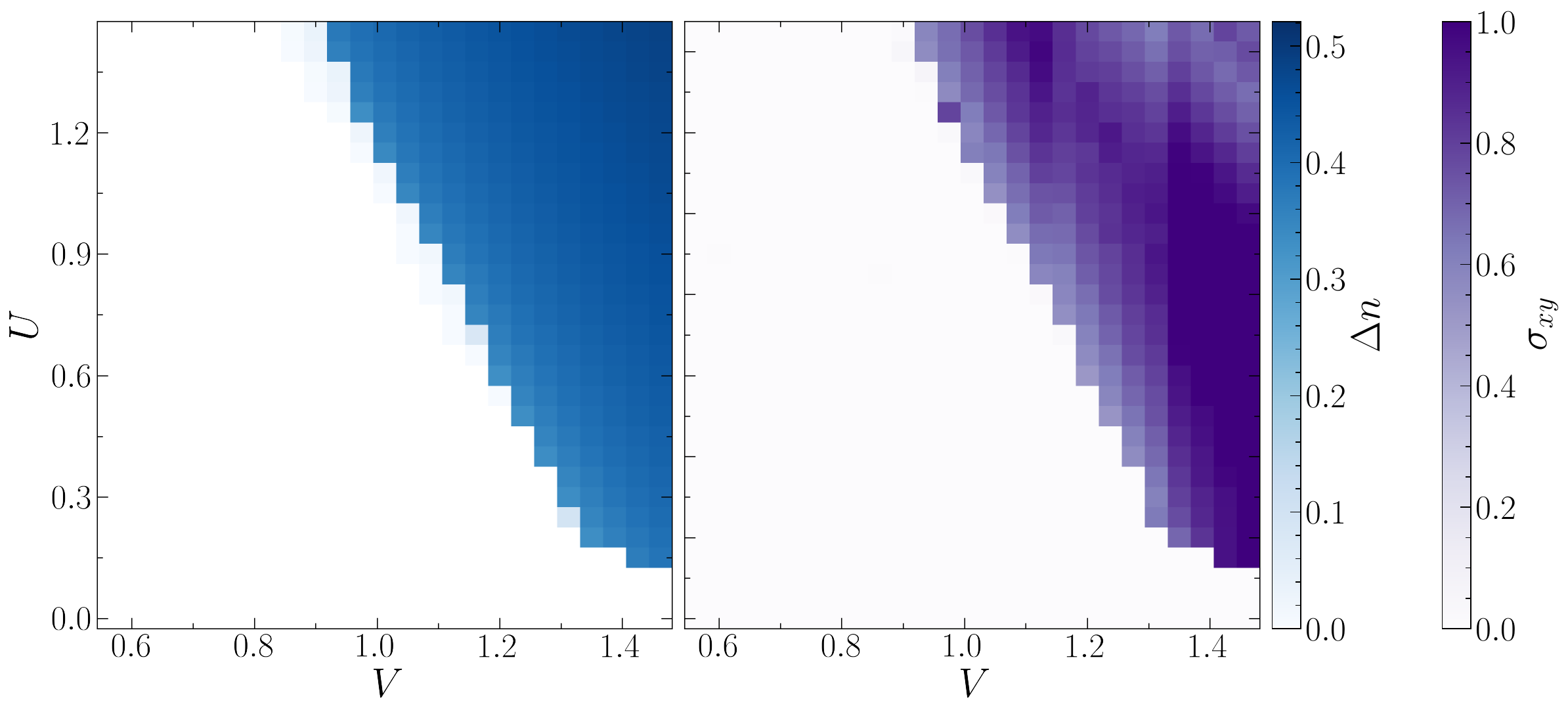}
     \caption{   
     The charge order $\Delta n (\bQ)$ and topological Hall response $\sigma_{xy}$ as a function of $U$ and $V$ for the band under discussion for $J_h/t = 2$.  Here, there is no weak-coupling limit at $\bQ = \boldsymbol{\Gamma}$. Rather there is a first-order transition to charge ordering at the SkX unit vector $\bQ = \bK/2$, with a real space representation in Fig.\ \ref{fig:charge_configuration}.}
    \label{phase_diagram_J-2}
\end{figure}

The full phase diagram as a function of $U,V$ is shown in Fig.\ \ref{phase_diagram_J-2}. At $U/t \sim 0$, the two layers are degenerate and there is no net Hall or polarization.  The effect of U contrasts with the competition at  $J_h/t = 4$.
We find that the inter-layer repulsion $U$ helps to stabilize the intralayer crystal formation for dispersive bands. 
The repulsion has a relative `squeezing' of the band-structure, which reduces the relative band-gaps. 
As the intralayer repulsion can be controlled with inter-layer distance, there is an intriguing amount of experimental control to induce such a transition.

\section{Discussion}
This work demonstrates several intriguing properties of skyrmion crystals and points to future directions. We demonstrate that there are flat bands with a propensity to form crystalline analogs of Landau levels within the spectrum of SkX.  We focus on a particular flat-band, just above half-filling, and explore the symmetry broken instabilities at partial filling of this degenerate band. \par

For large $J_h/t$, the polarization is suppressed by the large band-gaps, similar to the suppression of Landau level mixing for small interaction strength. For small $J_h/t$, there is the possibility of the nearest neighbor $V$ term driving band-mixings and reordering the charges.
In contrast, the large density of states of the flatter bands at large $J_h/t$ allows for weak-coupling instabilities to layer polarization without any single layer band-mixing. \par

A considerable amount of research has focused on fine-tuned or ``magic-angle" moir\'e configurations to realize flat bands  \cite{Andrei2020}.  There is however an increased realization that magnetic order can result in similar flat bands \cite{Paul2023}. 
This offers intriguing experimental directions to explore in magnetic bilayer systems. \par
Experimental fabrication could focus on hetrostructure device stacks. Of particular interest are the experimental realizations in \cite{Dohi2019, Legrand2020}. Additional realizations might occur in the triangular lattice Gd$_2$PdSi$_3$\cite{Kurumaji2019,  Hirschberger2020} with a suitable metallic layer or MoS$_2$/CrBr$_3$ \cite{Paul2023}.  
In particular, we emphasize for such experiments the importance of gating to consider the system just above half-filling of the entire crystal.
These fillings exhibit the most robust flat bands and topological response.
STM measurements would then be beneficial to determine the nature of any emergent correlated states \cite{Thompson2024, Liu2024,Zhang2024}. \par 

Future theoretical directions can focus on states within these bands with genuine topological order, which will require numerical analysis beyond the mean-field level.
Of particular interest in future studies is the required proximity to Landau level analogs for SkX flat bands to exhibit fractionalization \cite{Wang2021, Morales2024, Reddy2024b}.
SkX consisting of SU$(N>1)$ spinors or with topological charge $Q > 1$ may also in turn host bands with $C>1$. 
If these bands allow for FCI states, there may be many qualitative and quantitative differences to previous realizations \cite{Guerci2024}.

\par
Within Hartree-Fock however, these states already offer interesting broken-symmetry instabilities. We find at small $J_h$, intra-layer repulsion $V$ can form crystalline formation with nonzero Hall response.
One potential direction would be to explore what interactions would be necessary to realize ``anomalous Hall crystals" where the resulting crystalline state has an isolated band with non-zero Chern number $C$ \cite{dong2024_a,dong2023,  dong2024}. 
At larger $J_h$, combinations of interlayer repulsion $U$ and $V$ compete to drive an inversion-breaking polarization transition.
This phenomenology and competition can shed light on when polarized states are necessary to realize non-zero Hall effects \cite{Zhang2019, Repellin2020, Bultinck2020,Bultnick2020b, XieY2022, Crepel2023}. There are intriguing possibilities of coherent (rather than polarized) states that still exhibit an AHE \cite{Tao2024}.
Theoretical and experimental work on realizing such states in AF-SkX states would be enlightening. 
The ability to control these effects, through interlayer spacing for example, may offer control for topological and spintronics applications seeking to harness SkX states for quantum information storage and manipulation \cite{Wiesendanger2016, Jiang2017,Smejkal2018}.\par

\begin{acknowledgments}
    We thank D.\ Guerci, J.\ Huxford,  N.\ Paul, and A. Reddy for insightful discussions. 
	We acknowledge support from the Natural Sciences and Engineering Research Council (NSERC) of Canada. A.H. acknowledges support from a NSERC Graduate Fellowship (PGS-D). A.H. acknowledges support from a pre-doctoral fellowship at the Flatiron Institute while part of this work was being completed. The Flatiron Institute is a division of the Simons Foundation.  T.G. acknowledges support from the Center For Quantum Information and Quantum Control (CQIQC) Summer Fellowship. 
	Numerical computations were performed on the Niagara supercomputer at the SciNet HPC Consortium and the Digital Research Alliance of Canada. 
\end{acknowledgments}
% \bibliography{apssamp}% Produces the bibliography via BibTeX.
\bibliography{lib}

\appendix 
\onecolumngrid
\clearpage
\section{Skyrmion Classification}\label{appendix:classification}
Classification of skyrmion Crystal phases are determined by their homotopy group $\pi_3(\text{SO}(3))$. This topological charge is constructed from the combination of the skyrmion winding number and polarity $Q = P \cdot \mathcal{W}$. For a spin-texture given by \(\hat{\mathcal{S}}^{\ell}_{\a}(\br)\) where \(\a\in\{x, y, z\}\) and \(\ell\in\{1, 2\}\) represents the layer, we have 
\begin{equation}
    \label{skyrmion charge}
    Q^{\ell}=\frac{1}{8\pi}\dr\: \epsilon_{i j} \epsilon^{\a \b \g} \hat{{\cal S}}_\a^{\ell}(\br) \cdot\partial_i \hat{{\cal S}}_{\b}^{\ell}(\br) \cdot\partial_j \hat{{\cal S}}_{\g}^{\ell}(\br)\,.
\end{equation}
Switching to polar coordinates in real-space \(r, \phi\), and representing the spins on a Bloch sphere \(\hat{{\cal S}}^{\ell}(\br) = (\sin\Theta^{\ell}(\br)\cos\Phi^{\ell}(\br), \sin\Theta^{\ell}(\br)\sin\Phi^{\ell}(\br), \cos\Theta^{\ell}(\br))\), Eq.\eqref{skyrmion charge} simplifies to
\begin{equation}
\begin{split}
Q^\ell &= \frac{1}{4 \pi} \int_0^{\infty} d r \sin \Theta^{\ell} \frac{d \Theta^{\ell}}{d r} \cdot \int_0^{2 \pi} d \phi \frac{d \Phi^{\ell}}{d \phi} \\
&= \frac{\left[\hat{{\cal S}}_z^\ell(0)-\hat{{\cal S}}_z^\ell(\infty)\right]}{2} \frac{[\Phi^\ell(2 \pi)-\Phi^\ell(0)]}{2 \pi} = P^\ell \cdot \mathcal{W}^\ell
\end{split}
\end{equation}
The winding number can be further parameterized as $\Phi(\phi)=\nu \phi+\gamma$ where $\n$ is the vorticity and $\gamma$ is the helicity, characteristic to a given skyrmion texture. Furthermore, on a discrete lattice, the skyrmion charge is related to the chirality of the spin texture (typically set by the sign of the DM interaction which generates the SkX). \par 
We consider a unit cell from a skyrmion ansatz with a radius of $N_a$. The number of sites in the unit cell scales as $3N_a^2$. In this paper, we will focus on \(N_a=2\) which corresponds to a 12-site per layer unit cell as shown in Fig.\ \ref{fig:layers}. We have checked some results for $N_a = 3$ and do not see qualitative differences.
The ansatz we use is given as 

\begin{equation}
    \label{weiss}
    \hat{{\cal S}}^{\ell}(\br) = \begin{pmatrix}
    -\sin \left[ \pi \left(1 - \lceil r \rceil / 2\right)\right] y/r \\
    \sin \left[\pi \left(1 - \lceil r \rceil /2\right)\right] x/r \\
    \cos \left[\pi \left(1 - \lceil r \rceil/2\right) \right](-1)^{\ell-1}
\end{pmatrix}
\end{equation}
where $x\,,y\,,$ and \(r\) are measured with respect to the SkX unit-cell origin. Rotating the spins by \(\pi/2\) in plane switches from Bloch to Neel skyrmions.  Lastly, we are also agnostic to the helicity of the SkX phases (Neel or Bloch as shown in Fig.\ \ref{fig:layers}) because a spin-rotation can convert between the two and as such they do not affect band-structure results.
\par
In this paper, we consider a bilayer SkX model where the chirality is opposite between the two layers (which leads to opposite polarity) and hence we get opposite charges $Q^\ell = \pm 1$. We note that this is a generic form of $Q = \pm 1$, \textit{i.e.} AF-SkX pairs when the two layers are related by time-reversal followed by a spin-rotation around \(\hat{z}\), $\mathcal{T}\cdot R_{\hat{z}}(\pi)$ (effectively only flipping the \(z\) component of the spin). Symmetry-equivalent states can also be generated with only a $\mathcal{T}$ transformation (which flips the spin vector completely) or by the action of $\mathcal{T}\cdot R_{\hat{x}}(\pi)$ (which only flips the \(x\) component of the spin) that flip the vorticity and hence the winding number \(\mathcal{W}\). We caution that these AF-SkX states are distinct from skyrmion-anti-skyrmion pairs, as anti-vortex skyrmions break rotational symmetry.  
\section{Susceptibilities}\label{appendix:susceptibilities}
We calculate the susceptibility $\chi^0$ of the multi-sublattice model towards possible instabilities. The formulation is a direct extension of the simple single-band case. We also examine the effects of interactions on $\chi$ under the RPA approximation. For the bare susceptibility, we define the generalized density fields corresponding to our orderings as \(n_{\mathbf{R}, i}(\t) = c^{\dagger}_{\mathbf{R}, i, \s}(\t) c_{\mathbf{R}, i, \s}(\t)\), where \(\mathbf{R}\) refers to the unit-cell position, \(i\) refers to every degree of freedom other than spin (such as sublattice or layer), and the fermion spin \(\s\) is being summed over. In momentum space the corresponding vertex is given as
    \begin{equation}
        \label{charge vertex}
        n_{\boq, i}(i\W) = \d^{\s\s'}\int \frac{d^2k}{(2\p)^2}\int \frac{d\w}{(2\p)} c^{\dagger}_{\boq+\bok, i, \s}(i\W+i\w) c_{\bok, i, \s'}(i\w)\,,
    \end{equation}
    where \(\boq\) is the exchange momentum and \(\W\) is the exchange energy. 

Using these density fields, we  calculate the bare susceptibility of the model (discussed in details below). To get ordering tendencies of the model, we  diagonalize the zero-energy response matrix at all momenta. The eigenvector with the largest eigenvalue corresponds to possible orderings with the ordering vector being the momentum at which this maximum eigenvalue occurs. One can repeat this exercise after performing RPA, which can affect possible orderings in the model as interaction strength is slowly increased. Encountering a diverging eigenvalue of the response corresponds to a phase transition into a symmetry broken state, which should match qualitatively with mean-field results. Alternatively, one can also tune the temperature at strong coupling regime to extract information about phases beyond the first instability encountered at zero temperature when tuning the interaction.\par

The bare susceptibility, \(\c_{0, ij}(\mathbf{Q}, i\Omega) = \expval{n_{\mathbf{Q},i}(i\Omega)n_{\mathbf{Q}, j}^{\dagger}(i\Omega)}\) is equivalent to the usual density-response function. Diagrammatically, this corresponds to a generalized bubble diagram. For spin-rotation symmetric systems, the connected piece of the diagram is expressed as 
\begin{equation}
    \label{density suscep}
    \chi_{0, ij}(\mathbf{Q}, i\Omega) = -\int \frac{d^2k}{(2\p)^2}\int \frac{d\w}{(2\p)} G_{ij}(\mathbf{Q}+\mathbf{k}, i\Omega+i\omega)G_{ji}(\mathbf{k}, i\w)\,,
\end{equation}
where \(G_{ij}^{\a\b}(\mathbf{k}, i\w) = \d^{\a\b}G_{ij}(\mathbf{k}, i\w) = \expval{c_{i,\a}(\mathbf{k}, i\w) c^{\dagger}_{j, \b}(\mathbf{k}, i\w)}\) are the bare Green's functions calculated in mean-field theory. In terms of the quasi-particle dispersion, \(\e_n(\mathbf{k})\), and their corresponding wavefunctions \(U_{ij}(\mathbf{k})\), these Green's functions can be expressed as
\begin{equation}
    \label{greens function wavefunctions}
    G_{ij}(\mathbf{k}, i\w) = \sum_{n}U_{in}(\mathbf{k})\frac{1}{i\w - \e_n(\mathbf{k}) + \mu}U^{\dagger}_{nj}(\mathbf{k})\,.
\end{equation}
Substituting this expression back into \eqref{density suscep}, performing the Matsubara sum by hand, and analytically continuing the resulting expression \(i\Omega\rightarrow\Omega+i\eta\), we get
\begin{equation}
    \label{density suscep final}
    \chi_{0, ij}(\mathbf{Q}, \Omega) = \sum_{n,m}\int \frac{d^2k}{(2\p)^2} U_{jn}(\mathbf{Q}+\mathbf{k})U_{ni}^{\dagger}(\mathbf{Q}+\mathbf{k})U_{im}(\mathbf{k})U_{mj}^{\dagger}(\mathbf{k}) \frac{n_F(\e_{n}(\mathbf{Q}+\mathbf{k})-\mu) - n_F(\e_m(\mathbf{k})-\mu)}{\Omega + i\eta - (\e_{n}(\mathbf{k}+\mathbf{Q}) - \e_m(\mathbf{k}))}\,,
\end{equation}
where \(n_F\) is the Fermi distribution function.
\begin{figure}[ht]
 \centering
 \includegraphics[width=0.45\textwidth]{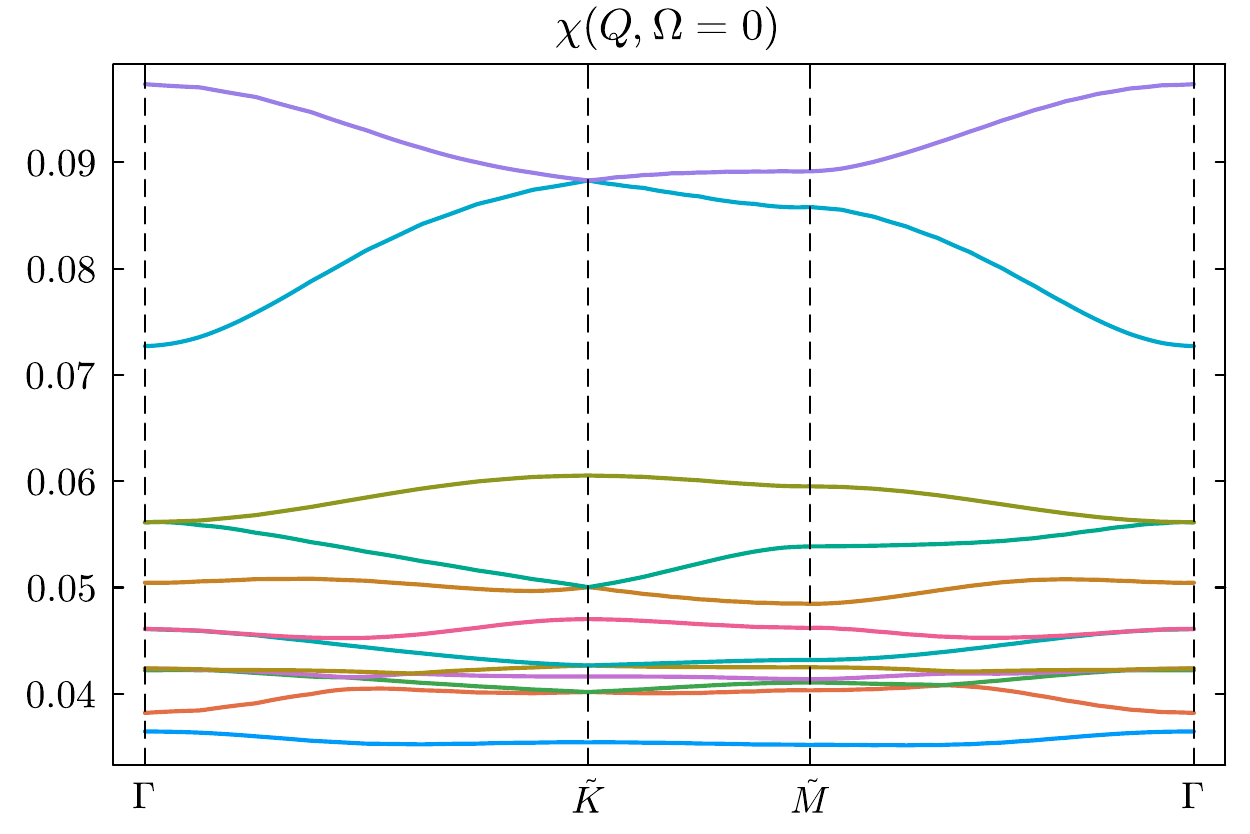}
 \includegraphics[width=0.45\textwidth]{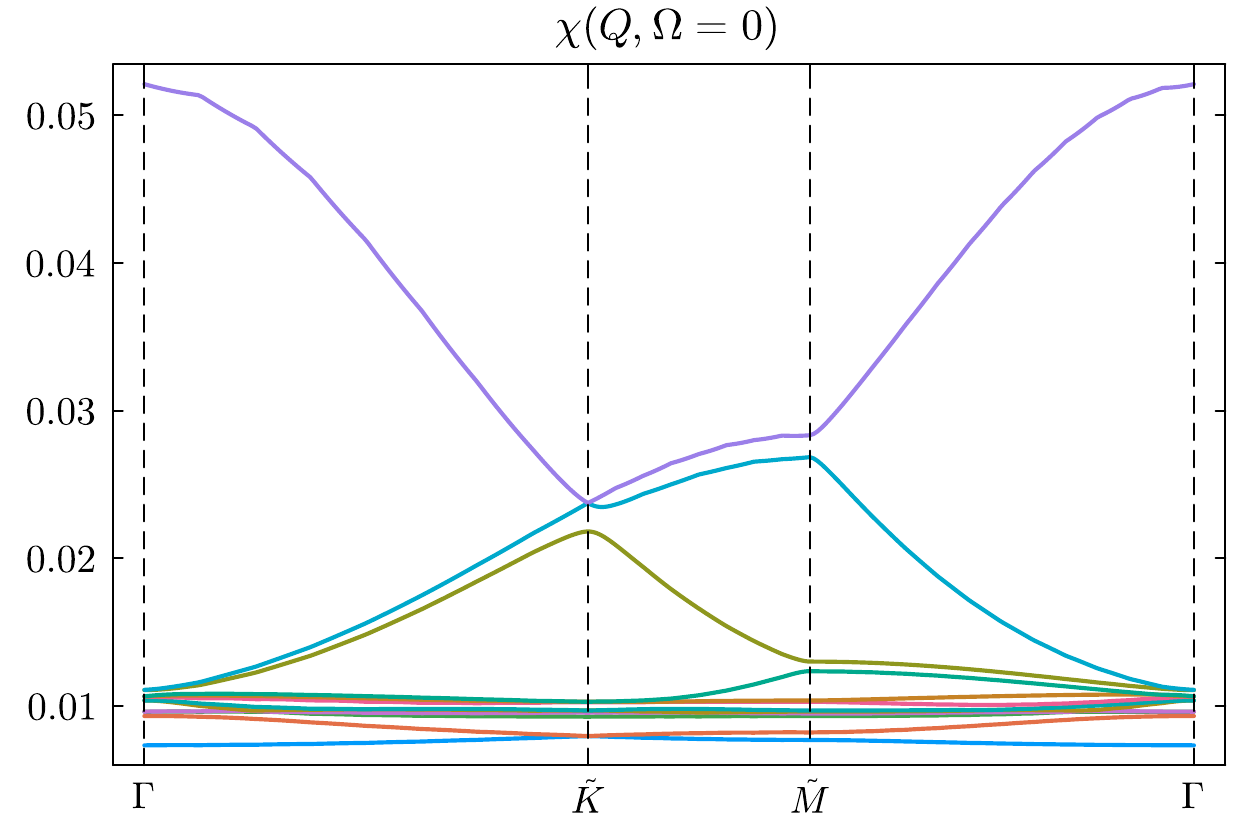}
 \caption{ The eigenvalues of the bare charge susceptibility matrix as defined in Eq.\ \ref{density suscep final} of the tight-binding model given in Eq.\ \ref{non-interacting} along a high-symmetry path in the SkX Brillouin Zone for (a) \(J_h/t = 2\), and (b) \(J_h/t=4\), at a filling \(\bar{n}= 25/48\) which sits at the half-filling of the flat band at \(J_h/t=4\).}
\label{fig:skyrmion bare suscep}
\end{figure}

\subsection{RPA}
To perform RPA, we decompose a general density-density interaction using Eq.\eqref{charge vertex} as
\begin{equation}
    \label{general heisenberg}
    \begin{split}
    \mh_{int} &= \sum_{\bod}\sum_{\bor, \t}V_{ij}(\bod) c^{\dagger}_{\bor, i, \s}(\t) c_{\bor, i, \s}(\t) c^{\dagger}_{\bor+\bod, j, \s'}(\t) c_{\mathbf{r}+\bod, j, \s'}(\t)\,,\\
        &= \sum_{\bod}\sum_{\bor, \t}V_{ij}(\bod)n_{\bor, i}(\t) n^{ \dagger}_{\bor + \bod, j}(\t)\\
        &= \sum_{\boq, \Omega}n_{\boq, i}(\Omega)\left[\sum_{\bod}V_{ij}(\bod)e^{-i \boq\vdot\bod}\right] n^{\dagger}_{\boq, j}(\Omega)\,.
    \end{split}
\end{equation}
Hence, under RPA, the susceptibility will have the form
\begin{equation}
    \label{magnetic RPA}
    \chi_{ij}(\boq, \W) = \left[\left(\mathds{1} - \chi_{0}(\boq, \Omega)\vdot \Tilde{V}(\boq)\right)^{-1}\vdot \chi_{0}(\boq, \Omega)\right]_{ij}\,,
\end{equation}
where the effective interaction matrix is \( \Tilde{V}_{ij}(\boq) = \sum_{\bod}V_{ij}(\bod)e^{-i \boq\vdot\bod}\), and \(\vdot\) denotes matrix multiplication in the sublattice/layer index \(i\).
\begin{figure}[ht]
 \centering
 \includegraphics[width=0.5\textwidth]{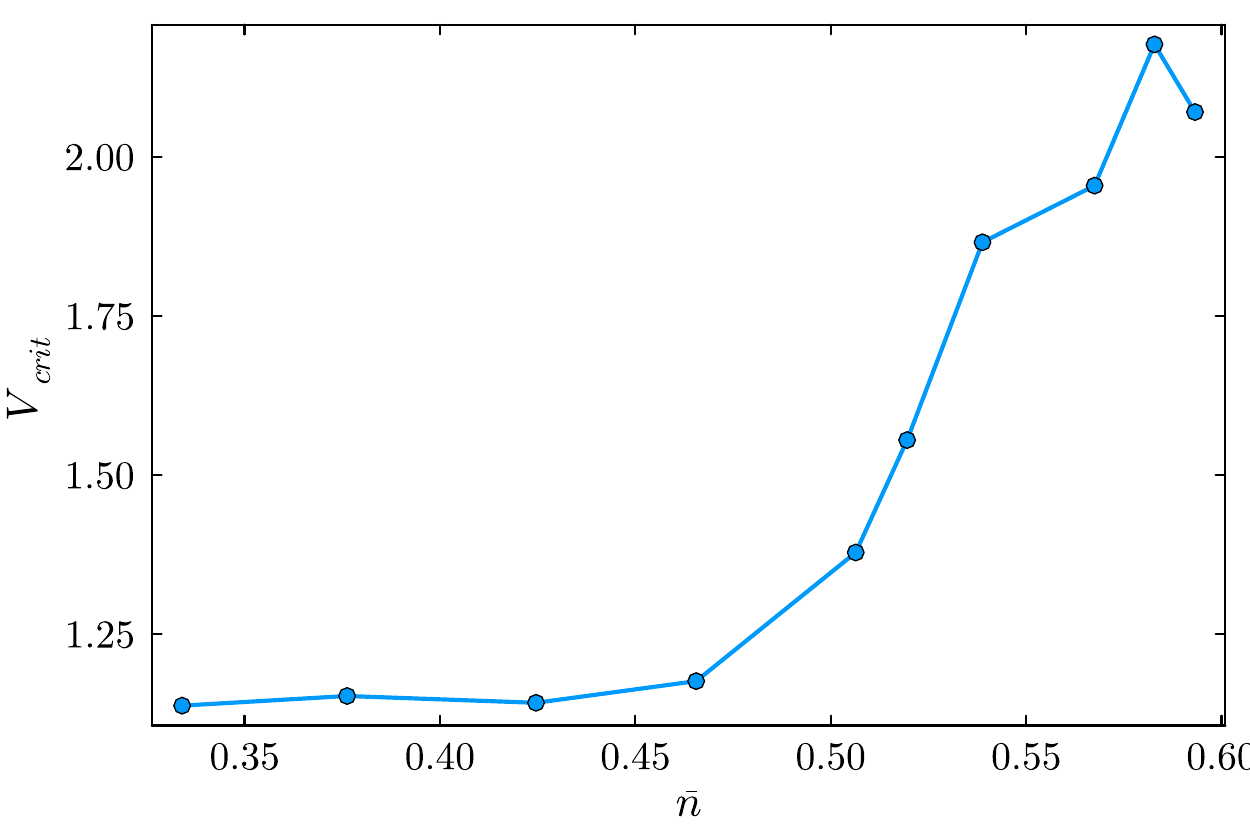}
 \caption{ The critical strength of the nearest neighbour density-density repulsion to induce a charge ordering at any \(\bQ\) as a function of the filling of the SkX model at \(J_h/t=2\). Incidentally, all of these instabilities are at \(\bQ=\boldsymbol{\Gamma}\), and hence these match exactly with our mean-field results assuming no translation symmetry breaking. Interestingly, repeating this exercise at \(J_h/t=4\) does not show any weak-coupling instabilities, even at the flat bands. The flat bands seem to not favour any one momenta to break translation symmetry along and hence we do not find any tendencies of charge ordering.}
\label{fig:V RPA}
\end{figure}

In our RPA analysis of the monolayer SkX model in the presence of nearest-neighbour density-density repulsion \(V\), we find contrasting behaviour at the two Hunds coupling being considered, \(J_h/t=2\) and \(J_h/t=4\). At \(J_h/t=2\), we find some instability to charge ordering at zero-momenta \(\bQ=0\), at some intermediate strength of the interaction as shown in Fig.\ \ref{fig:V RPA} as a function of filling \(\bar{n}\). This corroborates our momentum-space mean-field results which necessarily assumes lack of translation symmetry breaking. Meanwhile, at \(J_h/t=4\), we do \emph{not} find any instabilities at any momentum up to large values of the interaction, \(V/t>>1\).

\section{Charge Orders}\label{appendix:structure factor}
In this section we elaborate on the  different charge orders $n(\bQ),  \Delta n(\bQ)$ in detail, specifically in the two symmetry broken states that we find in the main text. \par

\begin{figure}[!ht]
 \centering
 \includegraphics[width=0.525\textwidth]{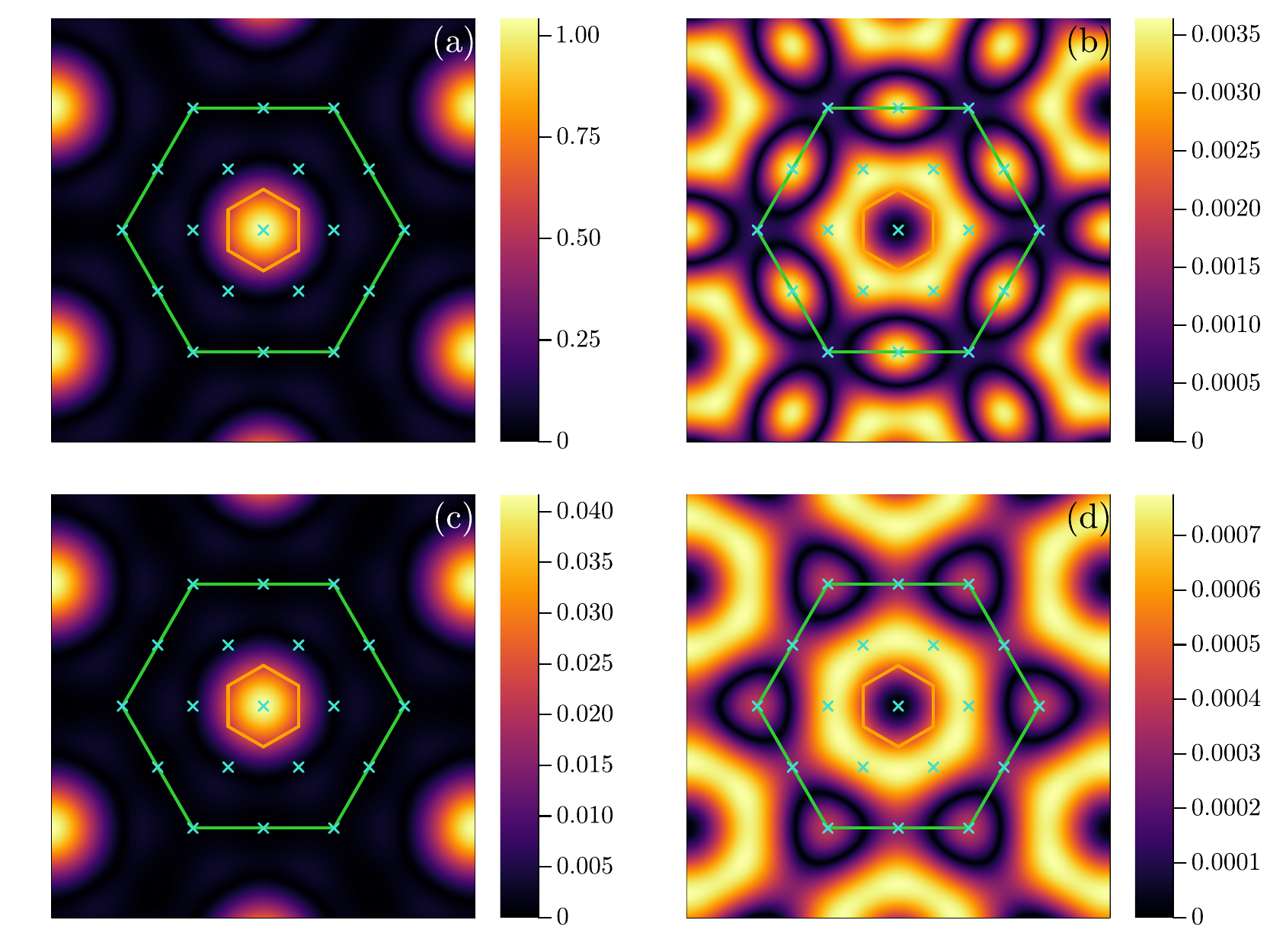}
 \caption{ The mean-field charge distribution in momentum space in the presence of the skyrmion with \(J_h/t=4\,,U/t=2\), and \(V/t=0\) at a filling of \(\bar{n}=25/48\). (a) The absolute value of the total charge density \(|n(\bQ)|\) in the triangular Brillouin zone. (b) The finite $q$ piece corresponding to part (a). (c) The absolute value of the polarization, \(|\D n(\bQ)|\), and (d) the uniform piece corresponding to (c). We see a net polarization between layers of \(\approx 0.04\), with a tiny modulation with strength \(\approx 0.001\) at some incommensurate momenta \(\bQ\neq 0\). The total charge density however looks very similar to that of the bare SkX model as shown in Fig.\ \ref{fig:skyrmion charges} }
\label{fig:skyrmion charges_J_4_U_2}
\end{figure}

The layer polarized case for the flat bands at $J_h = 4$ is shown in Fig.\ \ref{fig:skyrmion charges_J_4_U_2} where we find that (a) the total charge is mostly uniformly distributed with minor (\(\approx 0.003\)) modulations as shown in (b). Looking at the net polarization in (c), we again find that its mostly uniform (\(\approx 0.04\)) with negligible modulations in the full Brillouin zone. \par

\begin{figure}[t]
 \centering
 \includegraphics[width=0.525\textwidth]{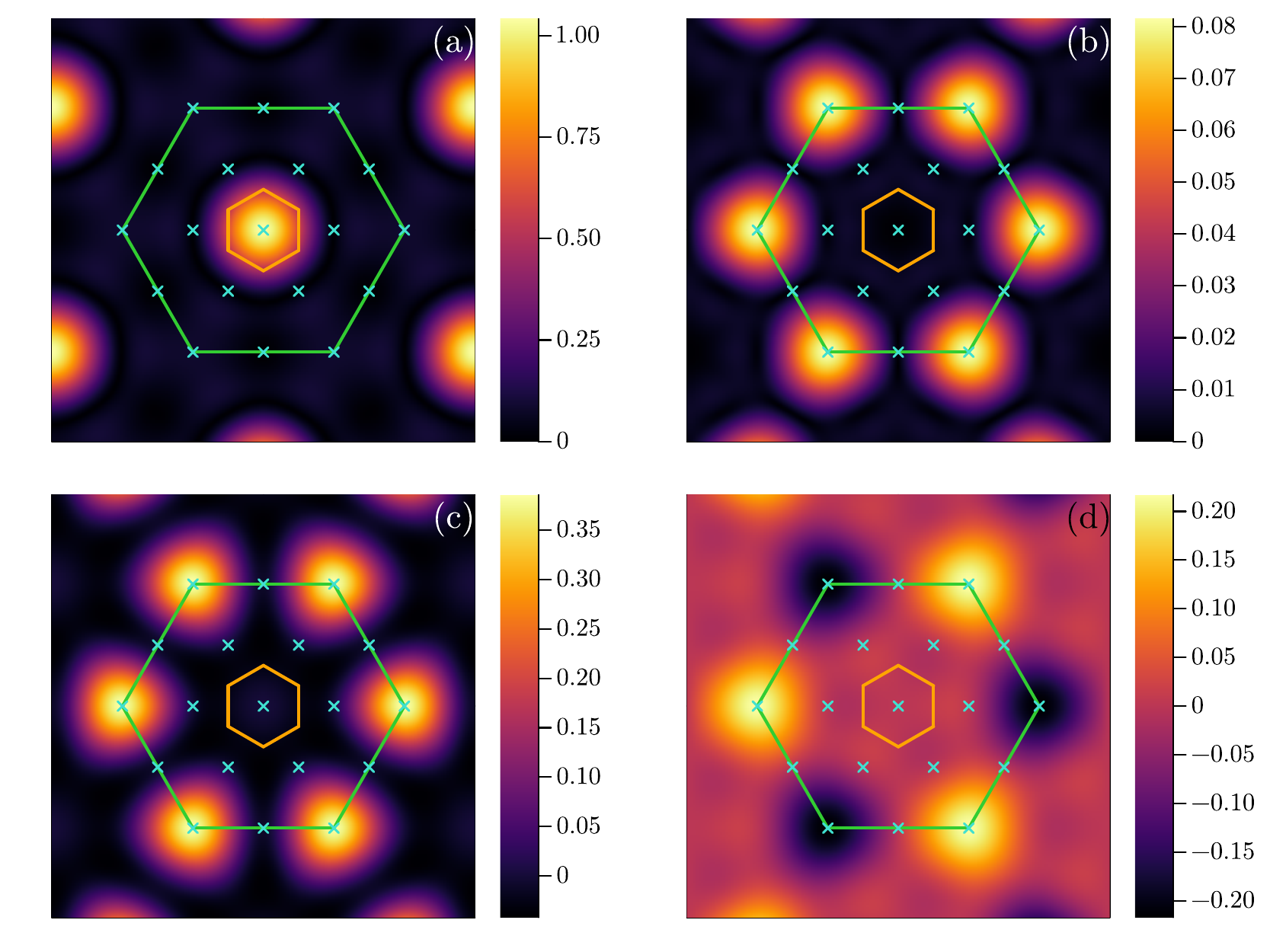}
The mean-field charge distribution in momentum space in the presence of the skyrmion with \(J_h/t=2\,,U/t=0.5\), and \(V/t=1.5\) at a filling of \(\bar{n}=25/48\). (a) The absolute value of the total charge density \(|n(\bQ)|\) in the triangular Brillouin zone. (b) The finite $q$ piece corresponding to part (a). (c) The real part of the finite $q$ piece of polarization, \(\Re\left(\D n(\bQ)\right)\), and (d) the imaginary part of the same. In this case, we find a vanishingly small net polarization of \(\approx 0.001\) and hence we do not plot the uniform and modulated polarizations separately. We instead find a polarization with strength \(\approx 0.4\) but at a non-zero momenta of \(\bQ=K, K'\). As is clear from the imaginary part of the polarization, this phase breaks \(\mathcal{R}\) from \(\mathcal{C}_6\) down to \(\mathcal{C}_3\).
\label{fig:skyrmion charges_J_2_V_1.5}
\end{figure}

In contrast, the intra-layer charge ordered state emerging from the dispersive bands at $J_h = 2$ shows drastic differences as plotted in Fig.\ \ref{fig:skyrmion charges_J_2_V_1.5}. 
Here we find that the finite $q$ component of the total charge $|n(\bQ)|$ has a sizable modulation (\(\approx 0.1\)) at the \(\bK\) points in the Brillouin zone, shown in $(b)$.
Looking at the modulated piece of the polarization, $\Delta n(\bQ)$ (both real (c) and imaginary (d)), we also find peaks at the same high-symmetry points. Lastly, the imaginary part clearly displays symmetry breaking of \(C_{6z} \rightarrow C_{3z}\).

\section{$J_h= 2$ Quantum Geometric Tensor}\label{appendix:J_2}
\begin{figure}[ht]
 \centering
 \includegraphics[width=0.99\textwidth]{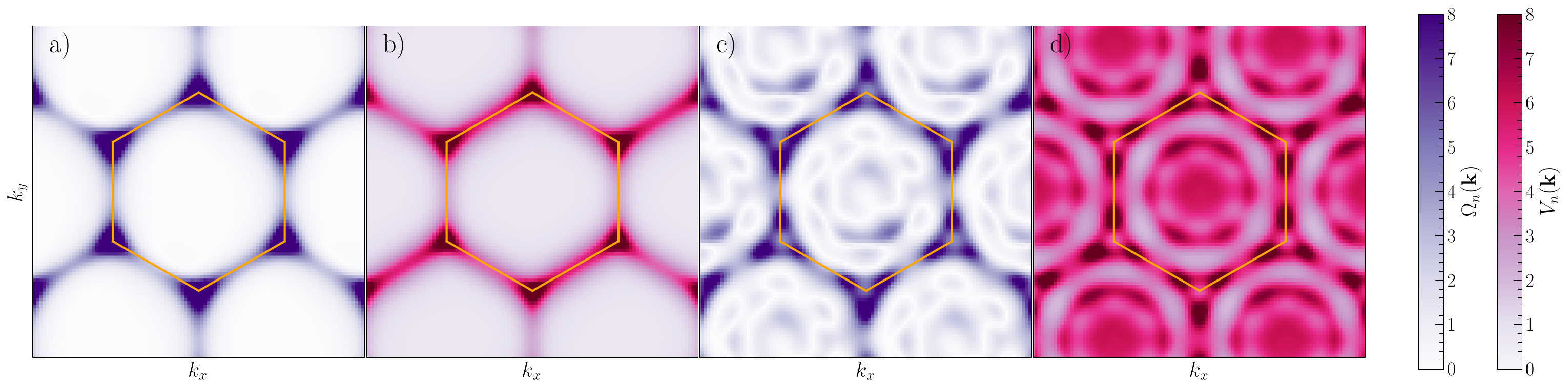}

 \caption{ (a) The momentum resolved  Berry curvature $\W_n(\bk)$, and (b) the quantum volume $V_n(\bk)$ for the first SkX band \(n=1\) at \(J_h/t=2\) which is dispersive. (c) The momentum resolved  Berry curvature $\W_n(\bk)$, and (d) the quantum volume $V_n(\bk)$ for the \(13^{\text{th}}\) SkX band at \(J_h/t=4\) which is also dispersive at \(J_h/t=2\). Both the bands now show structure as can be seen from peaks at high-symmetry points along the edge of the Brillouin zone.}
\label{fig:skyrmion metric Jh=2}
\end{figure}

In this section, we provide the The momentum resolved  Berry curvature $\W_n(\bk)$, and the quantum volume $V_n(\bk)$ for the first and thirteenth SkX band  at \(J_h/t=2\) as plotted in Fig. \ref{fig:skyrmion metric Jh=2}. The differences between the dispersive and approximate flat band are reduced because of the increased bandwidth of the the \(13^{\text{th}}\) SkX band.

\end{document}